\documentclass[twocolumn]{autart}    

\usepackage{amsmath, amssymb, amsfonts}  
\usepackage{graphicx} 

\usepackage{xcolor}

\usepackage{algorithm}
\usepackage{algpseudocode}

\usepackage{theorem}
\newtheorem{theorem}{Theorem}
\newtheorem{definition}{Definition}
\newtheorem{corollary}{Corollary}
\newtheorem{lemma}{Lemma}
\newtheorem{proposition}{Proposition}
\newtheorem{assumption}{Assumption}

\newtheorem{remark}{Remark}
\newtheorem{example}{Example}

\newcommand{\V}{\mathcal{V}}
\newcommand{\s}{\mathcal{S}}
\newcommand{\G}{\mathcal{G}}
\newcommand{\T}{\mathcal{T}}
\newcommand{\E}{\mathcal{E}}
\newcommand{\sop}{\mathrm{sp}}

\begin{document}
	\begin{frontmatter}
		\title{Adding links wisely: how an influencer seeks for leadership in opinion dynamics?}
		
		
		\author[KTH]{Lingfei Wang}\ead{lingfei@kth.se},    
		\author[KTH]{Yu Xing}\ead{yuxing2@kth.se},               
		\author[Sichuan]{Yuhao Yi}\ead{yuhaoyi@scu.edu.cn},  
		\author[Groningen]{Ming Cao}\ead{m.cao@rug.nl},
		\author[KTH]{Karl H. Johansson}\ead{kallej@kth.se}
		
		\address[KTH]{Division of Decision and Control Systems, School of Electrical Engineering and Computer Science, KTH Royal Institute of Technology, and Digital Futures, Stockholm, Sweden}  
		\address[Sichuan]{College of Computer Science, Sichuan University, and Institute of Clinical Pathology, West China Hospital, Sichuan University, Chengdu, China} 
		\address[Groningen]{ENTEG, Faculty of Science and Engineering, University of Groningen, Groningen, The Netherlands}        

		
		\begin{abstract}
			This paper investigates the problem of leadership development for an external influencer using the Friedkin–Johnsen (FJ) opinion dynamics model, where the influencer is modeled as a fully stubborn agent and leadership is quantified by social power. The influencer seeks to maximize her social power by strategically adding a limited number of links to regular agents. This optimization problem is shown to be equivalent to maximizing the absorbing probability to the influencer in an augmented Markov chain. The resulting objective function is both monotone and submodular, enabling the use of a greedy algorithm to compute an approximate solution. To handle large-scale networks efficiently, a random walk sampling over the Markov chain is employed to reduce computational complexity. Analytical characterizations of the solution are provided for both low and high stubbornness of regular agents. Specific network topologies are also examined: for complete graphs with rank-one weight matrices, the problem reduces to a hyperbolic 0–1 programmming problem, which is solvable in polynomial time; for symmetric ring graphs with circulant weight matrices and uniform agent stubbornness, the optimal strategy involves selecting agents that are sufficiently dispersed across the network. Numerical simulations are presented for illustration.
		\end{abstract}
	\end{frontmatter}

	\section{Introduction}
	
	Emergence of leadership is a prominent and complex phenomenon of human society, widely studied in sociological literature \cite{day2000leadership,moldoveanu2019future}. Among the various forms of leadership emerging in different social contexts, opinion leadership plays a crucial role, referring to an individual's capacity to influence others to adopt her viewpoint. Since individual opinions are dynamic and shaped by mutual interactions, understanding opinion leadership requires analysis within specific frameworks of opinion dynamics  \cite{degroot1974reaching,friedkin1990social,proskurnikov2015opinion,altafini2012consensus,ye2019influence}. Empirical studies have shown that leadership (or perceived charisma) is significantly influenced by an individual’s embedding within a social network, where nodes represent individuals and edges denote pairwise interactions \cite{carter2015social}. From this perspective, opinion leadership is often quantified by various network centrality measures \cite{borgatti2005centrality}. However, such measures are typically derived from the static structure of social networks, and fail to capture the dynamical nature of opinion formation. A more closely related concept is social power \cite{jia2015opinion,friedkin2016theory,tian2021social}, which is defined specifically for averaging-based opinion models. In particular, for an agent (i.e., individual) in the FJ model \cite{friedkin1990social}, whose opinion is updated as the weighted average of her neighbors' opinions and her own initial opinion, social power represents how much the initial opinion is weighted in the steady final opinions of the group that she belongs to. Other metrics have also been proposed for capturing opinion leadership in dynamic settings—for example, harmonic influence centrality \cite{vassio2014message,hunter2022optimizing}, which measures the average shift in group opinion when an agent's stance changes from one extreme to the other.
	
	The pursuit of opinion leadership is a natural driving force behind many social activities. For example, political leaders campaign to gain voter support, governments use mass media to persuade the public to take precautions during a pandemic, and negotiators in meetings employ strategic communication to convince others to accept their proposals. Reflecting these diverse contexts, various models have been proposed in the literature to capture how agents act to enhance their leadership. In \cite{catalano2024network}, a network formation game is explored where each agent aims to maximize her Bonacich centrality by strategically adding connections to the network. Coverage centrality is examined in \cite{medya2018group} as another measure of influence. In \cite{wang2024social} and \cite{wang2024maximizing}, the authors use the allocation of stubbornness in multiple FJ models as actions of the agents, and their payoffs are the social power in the opinion dynamics. Some of the results are validated by the data from the UN climate change conferences \cite{bernardo2023quantifying}. Additionally, \cite{ao2022agents} investigates a setting in which two external influencers compete to alter group opinions by adjusting the weights of their connections to other agents in the network. 
	
	Another related line of research is influence maximization, also known as seed selection in the literature \cite{kempe2003maximizing,morone2015influence}. In its classical form, an information propagation model is always considered, such as the independent cascade (IC) model and the linear threshold (LT) model \cite{aral2018social,chen2009efficient,li2018influence}. An external influencer initially selects some nodes as ``seeds" to inject information, then the information carried by the seeds eventually infects other nodes via social interactions, and the influencer's influence is measured by the number of final infected nodes. This formulation has also been adapted to opinion dynamics models, where the objective of the influencer is to steer the final opinions of the population toward desired values. In these settings, the influencer’s actions may include selecting a subset of agents, flipping their opinions, and rendering them fully stubborn—that is, their opinions remain fixed throughout the process—as considered in \cite{yildiz2013binary,gionis2013opinion,yi2021shifting,zhou2023opinion,raineri2025controlling}. Other approaches assume the influencer can modify the network by adding links to regular agents from fully stubborn agents that hold specific opinions \cite{hunter2022optimizing,zhu2025opinion}. Alternative optimization objectives have also been explored, such as minimizing opinion polarization within the network \cite{cinus2023rebalancing,tu2023adversaries}.
	
	In this paper, we adopt the FJ model as the opinion dynamics model. We consider an external influencer as a fully stubborn agent who participates in the opinion evolution by adding links to a selected subset of regular agents, thereby shaping the group’s final opinions. The influencer’s objective is to maximize her social power, a quantity that reflects her long-term influence and can be regarded as a form of harmonic influence centrality \cite{vassio2014message}. Unlike traditional influence maximization settings that focus on one-shot opinion shifts, our formulation emphasizes sustained leadership across potentially multiple FJ dynamics. To address the social power maximization problem, we conduct two complementary approaches. For general network topologies, we develop approximation algorithms to compute near-optimal solutions efficiently. For specific graph structures and agent stubbornness profiles, we derive analytical characterizations of the optimal strategy. Our main contributions can be summarized as follows:
	\begin{itemize}
		\item The FJ dynamics can be regarded as the state transition of a random walk over a Markov chain, referred to as a augmented Markov chain in this paper, with the initial opinions as absorbing states. It reveals that social power of each agent is equal to the absorbing probability associated to its initial opinion, and henceforth the problem for the influencer is to add links to maximize its own absorbing probability (Lemma \ref{le:equiv}).
		\item It proves that social power of the influencer, as a function of the agent set to be selected, is monotone and submodular (Propositions \ref{prop:monotone} and \ref{prop:submodular}). Therefore, a greedy algorithm can generate at least a near-optimal solution (Algorithm \ref{alg:1} and Theorem \ref{th:greedy}). To relieve some burden of computing the inverse of matrices in the greedy algorithm, especially for large networks, an algorithm that simulates random walks is proposed to approximate the social power of the influencer (Algorithm \ref{alg:2}). For the approximation algorithm, a high probability bound is proposed in Theorem~\ref{th:high_probability}.
		\item Homogenous stubbornness is considered. For this case, from the perspective of random walks over the augmented Markov chain, maximizing social power of the influencer is equivalent to minimize the hitting time to the corresponding absorbing state (Lemma \ref{le:equiv_markov}). This equivalence is further utilized to evaluate the influence of stubbornness and the weight of newly-added links on the optimization outcome (Propositions \ref{prop:equiv_single}, \ref{prop:big_stub} and Example~\ref{ex:homo_stub}).  
		\item For complete graphs with a rank-$1$ weight matrix, the influencer should choose agents that have the highest network centralities if the stubbornnes is homogenous, and should choose agents with the lowest stubbornness if their network centralities are the same (Corollary \ref{prop:rank1-special}). If both the stubbornness and network centralities are heterogeneous, the social power maximization problem becomes a constrained hyperbolic $0-1$ programming problem, which can be solved in polynomial time (Theorem \ref{th:complete}). On the other hand, if the graph is a ring, with its weight matrix as a symmetric circulant matrix, the influencer should choose two agents that have the largest distance from each other over the graph, given that the maximum number of agents to be selected is two (Theorem \ref{th:cycle}). Moreover, if multiple agents can be selected, it seems that a solution is to choose agents that are sufficiently dispersed over the graph (Example \ref{ex:ring_multi}).  
	\end{itemize}

	The paper is organized as follows: Section \ref{sec:preliminary} reviews some preliminary knowledge and formulates the problem; the results and algorithms for generic graph topologies are reported in Section \ref{sec:general}; the results for complete graphs and ring graphs are provided in Section \ref{sec:special}; Section \ref{sec:example} presents some numerical examples to illustrate the technical results; most of the proofs are shown in the Appendix. 
	
	A preliminary version of this paper was submitted to CDC25. That version only contains the results for generic graphs in this paper, with most of the proofs omitted. The results for specific graphs, which consist a significant part of this paper, appear for the first time.  
	
	\noindent\textbf{Notation.} All vectors are real column vectors and are denoted by bold lowercase letters $\bf{x},\bf{y},\dots$ The $i$-th entry of a vector $\mathbf{x}$ is denoted by $[\mathbf{x}]_i$ or, if no confusion arises, $x_i$. Matrices are denoted by the capital letters such as $A,B,\dots$, of entries $A_{ij}$ or $[A]_{ij}$. The identity matrix is denoted by $I_n$, with dimension sometimes omitted, depending on the context. The $n$-order vector and matrix with all entries being $0$ or $1$ are denoted by $\mathbf{0}_n$ or $\mathbf{1}_n$, respectively with the dimensions omitted if there is no confusion. Let $\mathbf{e}_i$ be the vector with the $i$th entry as $1$ and all the others as $0$. We use $[n]$ to represent the set $\{1,\dots,n\}$.
	Given a set $\mathcal{C}$, we use $|\mathcal{C}|$ to denote its cardinality. A square matrix $A$ is called (row) stochastic if $A\geq \mathbf{0}$ and $\mathbf{1}= A\mathbf{1}$. Given a real number $x$, let $[x]$ be the nearest integer to $x$. For two number sequences $f(n)$ and $g(n)$, we write $f(n)=O(g(n))$ if there exists a constant $C>0$ such that $|f(n)|<Cg(n)$ holds for all $n\in\mathbb{N}$. Given a random variable $X\in\mathbb{R}$, its expectation is denoted by $\mathbb{E}[X]$. Let $C(m_0,m_1,\dots,m_{n-1})$ denote the circulant matrix generated by the vector $(m_0,m_1,\dots,m_{n-1})^\top \in\mathbb{R}^n$, i.e.,
	\begin{equation*}
		\begin{aligned}
			C(m_0,&m_1,\dots,m_{n-1}):=\\
			&\left[\begin{array}{ccccc}
				m_0 & m_1 & m_2 & \cdots & m_{n-1} \\
				m_{n-1}  & m_0 & m_1 & \cdots & m_{n-2}\\
				m_{n-2} & m_{n-1} & m_0 & \cdots & m_{n-3}\\
				\vdots & \vdots & \vdots & \ddots & \vdots\\
				m_{1} & m_2 & m_3 & \cdots & m_0
			\end{array}\right].
		\end{aligned}
	\end{equation*}

	\section{Preliminaries and problem formulation}\label{sec:preliminary}
	
	\subsection{FJ model}
	
	Consider a network with nodes (agents) indexed in $\V=[n]$. It is represented by a directed graph $\mathcal{G}=(\mathcal{V},\mathcal{E},W)$, where $\mathcal{E}$ is a set of ordered pairs of nodes and $(i,j)\in\mathcal{E}$ represents the link from node $i$ to node $j$. The matrix $W$ is a stochastic weight matrix, such that for any $i,j\in\V$,  $(i, \, j ) \in \mathcal{E} $ if and only if $ w_{ji} >0$.  A (directed) {\em path} is a concatenation of directed links of $\mathcal{E}$. We say that node $i$ is connected to node $j$ if there is a directed path from $i$ to $j$. The graph $\mathcal{G}$ is called {\em strongly connected} if any two nodes are connected to each other.

	The FJ model is a DeGroot-like model for opinion dynamics in which some agents behave stubbornly, in the sense  that they defend their positions while discussing with the other agents \cite{friedkin1999influence}. The more stubborn an
	agent is, the lower is the total weight placed by the agent on others’ opinion.  If $n$ agents participate in a discussion, the FJ model is
	\begin{equation}\label{eq:FJ_original}
		\mathbf{x}(t+1)= (I -\Theta)W \mathbf{x}(t) + \Theta \mathbf{x}(0), \quad t=0,\, 1, \ldots
	\end{equation}
	where $\mathbf{x}(t)=(x_1(t),x_2(t),\dots,x_n(t))^\top\in\mathbb{R}^n$ is the opinion vector of the agents, $ W $ is a row-stochastic matrix, and $ \Theta = {\rm diag} (\theta_1, \ldots, \theta_n) $, with $ \theta_i \in [0,1]$ representing the stubbornness of agent $i$. 
	Stubbornness here means attachment of an agent to her own opinion, represented by the initial condition $ \mathbf{y}(0) $ at the beginning of the discussion ($ \theta_i =0 $ means agent $ i$ is not stubborn, $ \theta_i =1 $ means a fully stubborn agent).
	
	\begin{lemma}\label{le:1}\cite{proskurnikov2017tutorial}
		\label{lemma:DT1}
		Consider the FJ model \eqref{eq:FJ_original} over the graph $\G=(\V,\E,W)$. Assume $ \theta_i \in [0, 1]$ for all $i= 1, \ldots, n$. If for each $i\in\V$, either $ \theta_i >0$ or the agent $i$ is linked by a path from some agent $j$ with $\theta_j>0$, then 
		\begin{itemize}
			\item[(a)] $(I - \mathrm{\Theta}) W$ is Schur stable, i.e., $\rho ( (I - \Theta) W)<1$,
			\item[(b)]  The matrix  $P =(I - (I  - \mathrm{\Theta}) W)^{-1}\Theta $
			is stochastic,
			\item[(c)]  $\mathbf{x}(\infty)=\lim _{t\to +\infty} \mathbf{x}(t)= P\mathbf{y} (0)$.
		\end{itemize}
	\end{lemma}
	
	For an FJ model with convergent opinions, social power is defined as follows \cite{jia2015opinion,tian2021social}.
	
	\begin{definition}[Social power]\label{def:sp}
		Let the assumptions in Lemma~\ref{le:1} hold. The social power of each agent $i$ is $\sop_i:=\frac{1}{n}\mathbf{1}^\top P\mathbf{e}_i$.
	\end{definition}
	
	From Lemma \ref{le:1}, the solution matrix $P$ of an FJ model encodes the influence of each agent's initial opinion on the final opinion formation. Therefore, the social power of each agent $i$ is indeed the overall influence of agent $i$'s initial opinion on the group's final opinions. 
	
	\subsection{Problem formulation}
	
	Consider a groups of agents $\V=[n]$ over a graph $\G=(\V,\E,W)$, with $W$ as a stochastic matrix. The opinion evolution of the agents in $\V$ follows the FJ model \eqref{eq:FJ_original}. The following assumption is made throughout the paper.
	\begin{assumption}\label{assum:str_con}
		The graph $\G$ is strongly connected.
	\end{assumption}
	
	An external influencer, referred to as agent $0$ in this article, can add links to a subset of agents $\s\subset\V$. The agent $0$ is fully stubborn, i.e., $\theta_0=1$. The weights of the newly-added links are the same, that is, there exists $\omega\in[0,1)$ such that $w_{i0}=\omega$ holds for all $i\in\s$. Correspondingly, for each $i\in\s$, the weight of each incoming link $(j,i)\in\E$ becomes $(1-\omega)w_{ij}$. Let $\bar{\mathcal{G}}=(\bar{V},\bar{\E}, \bar{W})$ be the augmented graph by adding agent $0$ to the graph $\G$, with $\bar{V}=\V\cup \{0\}, \bar{\E}=\E\cup\{(0,i):i\in\s\cup\{0\}\}$ and 
	\begin{equation}
		\begin{aligned}
			\bar{W}
			= \left(\begin{array}{cc}
				1  & \mathbf{0} \\
				\omega\sum_{j\in\V}\mathbf{e}_j  &  (I-\omega\sum_{j\in\s}\mathbf{e}_j\mathbf{e}_j^\top)W
			\end{array} \right).
		\end{aligned}
	\end{equation}
	The opinion dynamics with the influencer then becomes
	\begin{equation}
		\bar{\mathbf{x}}(t+1)=(I_{n+1}-\bar{\Theta})\bar{W}\bar{\mathbf{x}}(t)+\bar{\Theta}\bar{\mathbf{x}}(0),
	\end{equation}
	with $\bar{\mathbf{x}}(t)=(x_0(t),x_1(t),\dots,x_n(t))^\top$ be the augmented opinion vector, and $\bar{\Theta}=\left(\begin{array}{cc}
		1 & \mathbf{0} \\
		\mathbf{0} & \Theta
	\end{array}\right)$.
	Under Assumption \ref{assum:str_con}, from Lemma \ref{le:1}, the opinions converge, and
	\begin{equation*}
		\begin{aligned}
			\bar{\mathbf{x}}(\infty)&=\lim_{t\to\infty}\bar{\mathbf{x}}(t)=\bar{P}\bar{\mathbf{x}}(0)=(I-(I-\bar{\Theta})\bar{W})^{-1}\bar{\Theta}\bar{\mathbf{x}}(0)\\
			&=
			\left(\begin{array}{cc}
				1 & 0 \\
				\mathbf{p}_0 & P 
			\end{array}\right)\bar{\mathbf{x}}(0),
		\end{aligned}
	\end{equation*}
	with
	\begin{equation}\label{eq:sol_FJ}
		\begin{aligned}
			\mathbf{p}_0&=(I-H(\s))^{-1}(I-\Theta)\omega\sum_{j\in\s}\mathbf{e}_j,\\
			P &=(I-H(\s))^{-1}\Theta, \\
			H(\s)&= \left(I-\omega\sum_{j\in\s}\mathbf{e}_j\mathbf{e}_j^\top\right)(I-\Theta)W.
		\end{aligned}
	\end{equation}
	The social power of agent $0$ is 
	\begin{equation}\label{eq:def_sop0}
		\mathrm{sp}_0=\frac{1}{n+1}\mathbf{1}^\top \mathbf{p}_0+\frac{1}{n+1}.
	\end{equation}
	
	The problem studied in this paper is stated as follows.
	\hfill\break
	
	\noindent\textbf{Problem}. Given the weight matrix $W$, the stubbornness matrix $\Theta$, and an integer budget $K\ge 1$, maximize the social power of agent~$0$ by adding links to no more than $K$ agents. Equivalently, solve the following optimization problem, with the set $\s$ collecting the linked nodes,  
	\begin{equation}\label{eq:op_follower}
		\begin{aligned}
			&\mathrm{Maximize}_{\s} \quad   \mathrm{sp}_0(\s)  \\
			&\quad\mathrm{s.t.} \quad \quad |\s|\leq K
		\end{aligned}
	\end{equation}
	Here we write $\mathrm{sp}_0$ as $\mathrm{sp}_0(\s)$ to show its dependence on $\s$. 
	
	\begin{remark}
		The external influencer, agent $0$, represents some opinion leader who insists her initial opinion, such as media outlets or companies promoting their products to consumers in the real world. Instead of directly altering the other agents' opinions, the influencer intentionally maximize the social power, which can be regarded as her long-term leadership in possibly multiple rounds of opinion formation processes that follow the FJ dynamics. 
	\end{remark}
	
	\begin{remark}\label{re:prob_compare}
		Mathematically, Problem \eqref{eq:op_follower} is to maximize the average opinion increase of the agents in $\V$ if agent $0$ flips her opinion from $0$ to $1$ while $x_i(0)=0,\forall i\in\V$, which is in fact harmonic influence centrality of agent $0$ \cite{vassio2014message,hunter2022optimizing}. Note that the problem of maximizing opinion sums by selecting followers and adding links to stubborn leaders are considered for the Degroot model in \cite{hunter2022optimizing,yildiz2013binary,zhou2023opinion,yi2021shifting}. Similar problems are also considered for the FJ model in \cite{gionis2013opinion,yi2021shifting,zhu2025opinion}, though with different assumptions: \cite{gionis2013opinion} studies the case of flipping the opinions of existing agents, \cite{yi2021shifting} examines continuous dynamics with partially stubborn influencers, while \cite{zhu2025opinion} assumes an undirected network structure, differing from the setting in this paper.  
	\end{remark}
	
	\begin{figure*}[t]
		\centering
		\includegraphics[trim=0cm 0cm 0cm 0cm,clip=true,width=13cm]{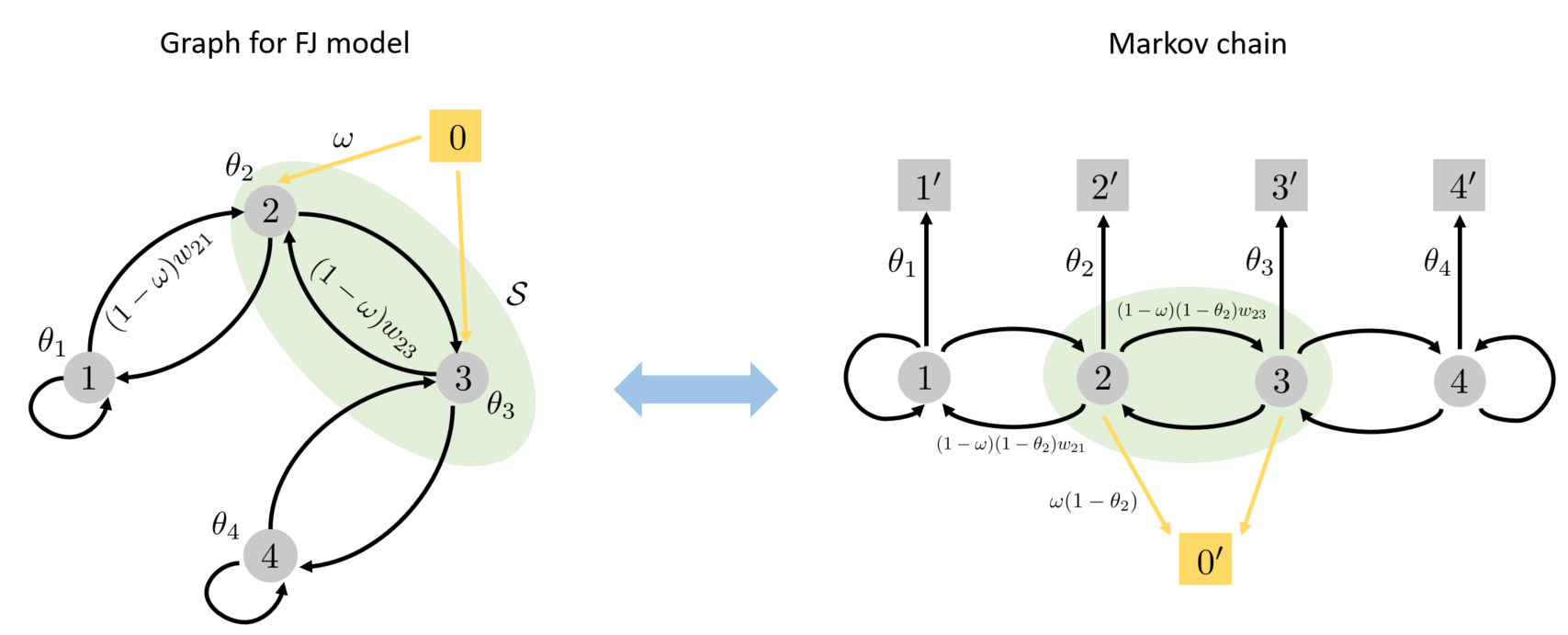}
		\caption{An FJ model with $4$ agents and external influencer $0$, together with its corresponding augmented Markov chain.}
		\label{fig:markov}
	\end{figure*}
	
	\section{Results for general graphs}\label{sec:general}
	
	In this section, we first show the connection of social power with the absorbing probability of a Markov chain. Secondly, we provide a greedy algorithm to compute an approximate solution of the problem \eqref{eq:op_follower}, and the Markov chain is utilized to reduce some computational complexity. Finally, for special cases of the agent stubbornness and link weight, we use the Markov chain to characterize the solution of \eqref{eq:op_follower} analytically.

	\subsection{Connection between social power and absorbing probability}\label{subsec:markov}
	
	For an FJ model, we construct an (absorbed) Markov chain, such that the social power of each agent is equal to the associated absorbing probability of a random walk generated by the Markov chain, which is called the augmented Markov chain of the FJ model in this paper.
	
	\begin{definition}[Augmented Markov chain]\label{def:MMC}
		Given an FJ model over $\mathcal{G}=(\V,\mathcal{E},W)$, with the matrix of stubbornness $\Theta$, its augmented Markov chain, denoted as $\mathcal{Y}(\G,\Theta)$, is a Markov chain $\{Y_t:t\in\mathbb{Z}_{\geq 0}\}$ with
		\begin{itemize}
			\item State space $\mathcal{T}\cup \mathcal{A}$, with $\mathcal{T}=\V$ as the transient states and $\mathcal{A}=\{1',2',\dots, n'\}$ as the absorbing states.
			\item Transition matrix $T=\left[\begin{array}{cc}
				(I-\Theta)W  & \Theta \\
				\mathbf{0} & I
			\end{array}\right]$, where for $i\in[n]$, the $i$-th column (row) corresponds to the transient state $i$, and the $(n+i)$-th state corresponds to the absorbing state $i'$.
		\end{itemize} 
	\end{definition}
	
	In Definition \ref{def:MMC}, each state $i'$ can be regarded as an auxiliary state of the state $i$, which corresponds to the initial opinion of agent $i$ in the FJ model. 
	
	For the FJ model with external influencer $0$, consider its augmented Markov chain $\mathcal{Y}(\bar{\mathcal{G}},\bar{\Theta})$ (denoted as $\bar{\mathcal{Y}}$ in the following if no confusion arise) over the augmented graph $\bar{\G}$. For the simplicity of notation, we merge the states $0$ and $0'$, and the transition matrix of $\bar{\mathcal{Y}}$ then becomes
	\begin{equation}\label{eq:trans_mat}
		\begin{aligned}
			T(\mathcal{S})=
			\left[\begin{array}{ccc}
				1  & \mathbf{0} & \mathbf{0} \\
				\omega(I-\Theta)\sum\limits_{j\in\mathcal{S}}\mathbf{e}_j  & H(\s) & \Theta \\
				\mathbf{0} & \mathbf{0} & I
			\end{array}\right],
		\end{aligned}
	\end{equation}
	where the first column (row) corresponds to the absorbing state $0'$; for $i\in[n]$, the $(i+1)$-th column (row) corresponds to the transient state $i$, and the $(n+i+1)$-th state corresponds to the absorbing state $i'$. The relationship between the FJ model and its augmented Markov chain is illustrated in Fig. \ref{fig:markov} for a $4$-node graph. 
	
	For the augmented Markov chain $\bar{\mathcal{Y}}=\{\bar{Y}_t:t\in\mathbb{Z}_{\geq 0}\}$, let the initial states uniformly distributed over the transient states and the absorbing state $0'$, i.e.,
	\[\mathbb{P}(\bar{Y}_0=0')=\mathbb{P}(\bar{Y}_0=i)=\frac{1}{n+1}, \quad \forall i\in[n].\]
	Let $\pi_i$ be the absorbing probability of the state $i'$ (here we let $0'=0$), i.e.,
	\[\pi_i:=\mathbb{P}(\bar{Y}_t=i' \text{ for some } t), \quad i=0, 1,\dots, n. \]
	Note that $(I-H(\s))^{-1}$ is the fundamental matrix of $\bar{\mathcal{Y}}$ \cite{kemeny1969finite}, and correspondingly,
	\noindent\begin{equation*}
		\begin{aligned}
			\pi_i&=\frac{1}{n+1}\mathbf{1}^\top \left(I-H(\s)\right)^{-1}\Theta\mathbf{e}_i\\
			&=\frac{1}{n+1}\mathbf{1}^\top P\mathbf{e}_i=\mathrm{sp}_i, \quad \forall i\in[n]
		\end{aligned}
	\end{equation*}
	and $\pi_0=\sop_0$.
	Therefore, the following lemma can be directly obtained.
	
	\begin{lemma}\label{le:equiv}
		Consider the FJ model over the graph $\G$. Let Assumption \ref{assum:str_con} hold. The optimization problem \eqref{eq:op_follower} is equivalent to the following problem 
		\begin{equation}\label{prob:markov}
			\mathrm{Maximize}_{|\s|\leq K} \quad   \pi_0(\s),  
		\end{equation}
		where $\pi_0(\s)$ is the absorbing probability to $0'$ for the augmented Markov chain $\bar{\mathcal{Y}}$.  
	\end{lemma}
	
	As mentioned in Remark \ref{re:prob_compare}, social power is a measurement of centrality. Lemma \ref{le:equiv} provides an equivalent characterization of the centrality by using absorbing probabilities of a Markov chain. This also paves the way for the algorithm design and analysis in the following subsections. Note that in \cite{gionis2013opinion}, a weighted sum of such absorbing probabilities is also shown to be equal to the sum of final opinions in the FJ model.
	
	\subsection{Algorithm design}\label{sec:alg}
	
	This subsection will provide an algorithm that generates an approximate solution to the problem \eqref{eq:op_follower}. 
	
	We write $P$ and $\mathbf{p}_0$ as $P(\s)$ and $\mathbf{p}_0(\s)$ to emphasize their dependence on $\s$. The following propositions prove that $\mathrm{sp}_0(\s)$ is monotonically increasing and submodular with $\s$, respectively. 
	
	\begin{proposition}\label{prop:monotone}
		Let Assumption \ref{assum:str_con} hold. For any $\s\subset \V$ and $i\notin \s$, it holds $\sop_0(\s\cup \{i\})>\sop_0(\s)$.
	\end{proposition}
	
	\begin{proposition}\label{prop:submodular}
		Let Assumption \ref{assum:str_con} hold. The social power $\sop_0$ is a submodular function of $\s$.
	\end{proposition}
	
	The proof of Propositions \ref{prop:monotone} and \ref{prop:submodular} are in Appendix \ref{ap:1}.
	
	With Propositions \ref{prop:monotone} and \ref{prop:submodular}, a greedy algorithm can be designed to find an approximated solution to the problem \eqref{eq:op_follower}, which is given as Algorithm \ref{alg:1}. The following Theorem \ref{th:greedy} is then obtained as a direct application of well-known properties of monotone submodular functions \cite{nemhauser1978analysis}. 
	
	\begin{algorithm}[H]
		\caption{Greedy Algorithm for Social Power Maximization}\label{alg:1}
		\hspace*{\algorithmicindent} \textbf{Input:} A strongly connect graph $\G=(\V, \mathcal{E}, W)$; maximum number of new links $K$; a diagonal matrix of stubbornness $\Theta$; link weight $\omega$ \\
		\hspace*{\algorithmicindent} \textbf{Output:} A subset of nodes $\s\subset\V$ with $|\s|=K$
		\begin{algorithmic}[1]
			\State Initialization: $\s\gets \emptyset, \s^c\gets \V\backslash \s$
			\For{$k= 1$ to $K$}
			\State $\sop_\text{max}\gets 0, i_\text{max}\gets 0$
			\For{$i=1$ to $n$}
			\State Compute $\sop_0(\s\cup \{i\}) $ by using Eqs. \eqref{eq:sol_FJ} and \eqref{eq:def_sop0}
			\If{$\sop_0(\s\cup \{i\})> \sop_\text{max}$}
			\State $\sop_\text{max}\gets \sop_0(\s\cup \{i\})$, $i_\text{max}\gets i$
			\EndIf
			\EndFor
			\State Update: $\s\gets \s\cup\{i_\text{max}\}$
			\EndFor
		\end{algorithmic}
	\end{algorithm}
	
	\begin{theorem}\label{th:greedy}
		Consider the problem \eqref{eq:op_follower} for the graph $\mathcal{G}$. Let $\s^\ast$ be a solution and $\s^\text{gd}$ be the output of Algorithm \ref{alg:1}. It then holds
		\[\sop_0(\s^\text{gd})\geq (1-\frac{1}{e})\sop_0(\s^\ast), \]
		where $e$ is the natural logarithm.
	\end{theorem}
	
	Note that in Algorithm \ref{alg:1}, the computation of each $\sop_0(\s\cup\{i\})$ (in Step 5) requires the inverse of $I-(I-\omega\sum_{j\in\s}\mathbf{e}_j\mathbf{e}_j^\top)(I-\Theta)W$, for which the complexity is $O(n^3)$. Therefore, the computational complexity of Algorithm \ref{alg:1} is $O(K(n-\frac{K-1}{2})n^3)$. The computation of matrix inverse can be impractical when the dimension $n$ is large. To reduce computational complexity, we can make use of random walks generated by the augmented Markov chain corresponding to the FJ model, and compute an approximation of $\sop_0(\s)$ by the following algorithm.
	
	\begin{algorithm}[H]\label{alg:2}
		\caption{Approximation of $\sop_0(\s)$}\label{alg:2}
		\hspace*{\algorithmicindent} \textbf{Input:} $\G=(\V, \mathcal{E}, W)$; $\s$; $\Theta$; $\omega$; number and length of random walks $r, \ell$ \\
		\hspace*{\algorithmicindent} \textbf{Output:} An approximation of the absorbing probability of the state $0$, $\sop_0^\text{apx}(\s)$
		\begin{algorithmic}[1]
			\State Initialization: $\sop_0^\text{apx}(\s)\gets 0$
			\For{$k= 1$ to $r$}
			\State Generate a random walk $\mathcal{Y}^{(k)}$ from the transition matrix \eqref{eq:trans_mat} with length no larger than $\ell$, and the starting state is uniformly distributed over $[n]$
			\If{the $\mathcal{Y}^{(k)}$ is absorbed by $0$}
			\State $\sop_0^\text{apx}(\s)\gets \sop_0^\text{apx}(\s)+1$
			\EndIf
			\EndFor
			\State Update: $\sop_0^\text{apx}(\s)\gets \sop_0^\text{apx}(\s)/r$
		\end{algorithmic}
	\end{algorithm}
	
	For Algorithm \ref{alg:2}, a natural question is how large $r$ and $\ell$ are required to guarantee that $\sop_0^\text{apx}(\s)$ is close to $\sop_0(\mathcal{S})$. For the cases that all the agents have positive stubbornness, a high probability bound of $|\sop_0^\text{apx}(\s)-\sop_0(\mathcal{S})|$ can be given as follows, in which 
	\[\sop_0^\ell(\s):=\frac{1}{n}\mathbf{1}^\top \sum_{k=0}^\ell H(\s)^k(I-\Theta)\omega\sum_{j\in\s}\mathbf{e}_j<\sop_0(\s). \]
	
	\begin{theorem}\label{th:high_probability}
		Let Assumption \ref{assum:str_con} hold. Suppose that $\theta_i\geq\theta>0$ hold for all $i\in\V$. In Algorithm \ref{alg:2}, given $\epsilon, \delta\in(0,1)$ and $\sigma\in(0,1)$, let 
		\begin{equation*}
			\begin{aligned}
				r &\geq \frac{3}{\sigma^2 \epsilon^2 \sop_0^\ell(\s)}  \log (\frac{2}{\delta}), \\ 
				\ell&\geq \frac{\log(\theta(1-\sigma)\epsilon \sop_0^\ell(\s))-\log \omega}{\log(1-\theta)}-2.
			\end{aligned}
		\end{equation*}
		Then, for any $\s$ with $|\s|\geq 1$, it holds
		\[ \mathbb{P}(|\sop_0^\text{apx}(\s)-\sop_0(\mathcal{S})|< \epsilon \sop_0^\ell(\s))\geq 1-\delta. \]
	\end{theorem}
	
	The proof of Theorem \ref{th:high_probability} is given in Appendix \ref{ap:1}.
	
	\begin{remark}
		From \eqref{eq:sp_truncate} in the proof, it can be seen that $\sop_0^\ell(\s)\geq \frac{\omega(1-\theta_{\max})}{n}$, where $\theta_{\max}:=\max_{i\in\V}\{\theta_i|\theta_i<1\}<1$ (we only need to consider that for any $i\in\s$, it holds $\theta_i<1$). This means that $r= O(n)$ and $\ell = O(\log(n))$. Moreover, the complexity of one-step random walk is $O(d_{\max})$, with $d_{\max}$ the maximum degree of graph $\G$. Therefore, the complexity of Algorithm \ref{alg:2} is $O(n^2\log(n))$, which indicates that if $n$ is large, Algorithm \ref{alg:2} can reduce the computational complexity and obtain value of $\sop_0(\s)$ in a relatively high accuracy.  
	\end{remark}

	\subsection{Influence of stubbornness and link weight $\omega$}\label{subsec:homo_stub}
	In this subsection we directly consider the problem \eqref{prob:markov}, under the assumption that all the agents have homogenous stubbornness, i.e., for some $\theta\in(0,1)$ it holds that $\theta_i=\theta, \forall i\in\V$. For this case, the problem \eqref{prob:markov} is in fact to minimize hitting time of the augmented Markov chain to the absorbing states. 
	
	\begin{lemma}\label{le:equiv_markov}
		For the augmented Markov chain $\bar{\mathcal{Y}}=\{\bar{Y}_t:t\in\mathbb{Z}_{\geq 0}\}$ with $\theta_i=\theta\in(0,1), \forall i\in\V$ and $\mathbb{P}(\bar{Y}_0=i)=\frac{1}{n}, \forall i\in\V$, define $\bar{\tau}=\min_{t\geq 0} \{t: \bar{Y}_t\in \mathcal{A}\}$. The optimization problem \eqref{prob:markov} is equivalent to 
		\begin{equation}\label{prob:hitting_time}
			\mathrm{Minimize}_{|\s|= K} \quad \mathbb{E}_{\bar{\mathcal{Y}}}[\bar{\tau}],
		\end{equation}
		where the subscript is to specify the Markov chain.
	\end{lemma}
	
	\noindent\textbf{Proof}. Define 
	\begin{equation}\label{eq:def_f}
		f(\s;\theta,\omega):= \frac{1}{n}\mathbf{1}^\top \big(I-(1-\theta)(I-\omega\sum_{j\in\s}\mathbf{e}_j\mathbf{e}_j^\top) W\big)^{-1}\mathbf{1}.
	\end{equation}
	According to the monotonicity of the cost function and that $\mathbf{1}^\top \mathbf{p}_0+\mathbf{1}^\top P\mathbf{1}=n$, the problem \eqref{prob:markov} then becomes 
	\begin{equation}
		\mathrm{Minimize}_{|\s|= K} \quad   \frac{n\theta}{n+1}f(\s;\theta,\omega)+\frac{1}{n+1},
	\end{equation}
	that is
	\begin{equation}\label{prob:homo_stub}
		\mathrm{Minimize}_{|\s|= K} \quad   f(\s;\theta,\omega).
	\end{equation}
	On the other hand, $f(\s;\theta,\omega)$ is the expected number of times that the Markov chain $\bar{\mathcal{Y}}$ visits the transient states $\V$ \cite{chen2008expected} before being absorbed, which is equal to the expected hitting time of $\bar{\mathcal{Y}}$ to the absorbing states $\mathcal{A}$ (i.e., $\{0',1',\dots,n'\}$), i.e., $\mathbb{E}_{\bar{\mathcal{Y}}}[\bar{\tau}]$. The desired conclusion then holds. \hfill$\square$

	\begin{remark}
		In the literature \cite{adriaens2023minimizing} and \cite{haddadan2021repbublik}, similar problems of reducing hitting time for absorbing Markov chains are studied, under the scenario of reducing the so-called ``structural bias" of networks. Compared to the literature, the problem \eqref{prob:hitting_time} considers a more general network structure, in the sense that the transition probabilities from each node (or state) to its neighbors do not need to be equal. Moreover, the parameters $\theta$ and $\omega$, not considered in \cite{adriaens2023minimizing} and \cite{haddadan2021repbublik}, can affect the solution of the optimization problem \eqref{prob:markov} (or \eqref{prob:hitting_time}), as discussed in what follows.  
	\end{remark}
	
	As seen in \eqref{eq:def_f}, the value of $f(\s;\theta,\omega)$ depends on $\theta$ and $\omega$ nonlinearly due to the matrix inverse, which makes it hard to disentangle the influence of the parameters on the optimal selection of $\s$. Nevertheless, for specific range of the parameters, we can still obtain some intuitive conclusions. Specifically, in the remaining part of this subsection, we make a further assumption that $K=1$, i.e., only one agent is selected by agent $0$ to which a link is added. For this case, any $f(\{i\};\theta,\omega)$ is simply written as $f(i;\theta,\omega)$.
	
	We at first consider the case that $\theta$ is small, as specified by the following assumption.
	
	\begin{assumption}\label{assum:small_stub}
		Given $\omega>0$, $\theta$ is small enough such that $f(i;\theta,\omega)-f(i_\ast;\theta,\omega)>0$ holds for all $i\not= i_\ast$, where $i_\ast=\mathop{\arg\min}_{j\in\V} f(j;0,\omega)$.
	\end{assumption}
	
	\noindent Assumption \ref{assum:small_stub} requires that $i_\ast$ is the only solution to the problem $\mathrm{Minimize}_{i\in\V} f(i;0,\omega)$. As $f(i;\theta,\omega)$ is continuous with $\theta$, there exists $\tilde{\theta}\in(0,1)$ such that for any $\theta<\tilde{\theta}$, Assumption~\ref{assum:small_stub} holds.
	
	For the augmented Markov chain $\mathcal{Y}(\G, \mathbf{0})=\{Y_t:t\in\mathbb{Z}_{\geq 0}\}$ of the FJ model without agent $0$, define
	\[\tau_{ji}=\min\{t>0: Y_t=i|Y_0=j\} \]
	as the first time that the Markov chain hits $i\in\V$, given that the initial state is $j\in\V$. 
	
	
	
	\begin{proposition}\label{prop:equiv_single}
		Consider the FJ model over the graph $\G$ with $\theta_i=\theta,\forall i\in\V$. Let Assumptions \ref{assum:str_con} and \ref{assum:small_stub} hold. The optimization problem \eqref{prob:markov} with $K=1$ is equivalent to 
		\begin{equation}\label{prob:single}
			\mathrm{Minimize}_{i} \quad   \frac{1}{n}\sum_{j\not= i}\mathbb{E}_{\mathcal{Y}(\G, \mathbf{0})}[\tau_{ji}]+\frac{1-\omega}{\omega}\mathbb{E}_{\mathcal{Y}(\G, \mathbf{0})}[\tau_{ii}]
		\end{equation}
	\end{proposition}
	
	The proof of Proposition \ref{prop:equiv_single} is given in Appendix \ref{ap:equiv_markov}. 
	
	\begin{remark}\label{re:small_stub}
		The cost function of \eqref{prob:single} is a weighted combination of two terms: the first term $\frac{1}{n}\sum_{j\not= i}\mathbb{E}_{\mathcal{Y}(\G, \mathbf{0})}[\tau_{ji}]$ represents the average expected time for a random walk from the other agents to reach $i$, and the second term $\mathbb{E}_{\mathcal{Y}(\G, \mathbf{0})}[\tau_{ii}]$ is the expected time from $i$ to come back. Both terms only depend on the graph $\G$. For the problem \eqref{prob:single} with a large $\omega$ (i.e., close to $1$), it is more important to minimize the first term, which can be regarded as a measurement of the overall reachability to $i$ in the graph; on the other hand, if $\omega$ is small (i.e., close to $0$), the local structure of each agent in the graph affects more on the solution of \eqref{prob:single}. This is illustrated by Example \ref{ex:homo_stub} in Section \ref{sec:example}.
	\end{remark}
	
	Now consider the case that $\theta$ is close to $1$. Denote
	\[g_i:=(1-\omega)W_{ii}+\sum_{j\not= i}W_{ji}. \]
	We make the following assumption. 
	
	\begin{assumption}\label{assum:big_stub}
		There exists only one agent $i^\ast$ such that $g_{i^\ast}=\max_{i\in\V}g_i$. Moreover, for $\delta_g:=\min_{i\in\V}\{g_{i^\ast}-g_i:i\not=i^\ast\}$, it holds $\theta>\frac{n}{\delta_g+n}$.
	\end{assumption}
	
	\begin{proposition}\label{prop:big_stub}
		Consider the FJ model over the graph $\G$ with $\theta_i=\theta,\forall i\in\V$. Let Assumptions \ref{assum:str_con} and \ref{assum:big_stub} hold. For the optimization problem \eqref{prob:markov} with $K=1$, the solution is $\s^\ast=\{i^\ast\}$.
	\end{proposition}
	
	The proof of Proposition \ref{prop:big_stub} is given in Appendix \ref{ap:equiv_markov}. From Proposition \ref{prop:big_stub}, it can be seen that if $\theta$ is close to $1$, agent $0$ will choose an agent $i$ with the largest weighted sum of the weights on all the outgoing links, with $1-\omega$ as the weight assigned to $W_{ii}$. 
	
	
	\begin{remark}
		Combining Propositions \ref{prop:equiv_single} and \ref{prop:big_stub}, it can be seen that for small stubbornness $\theta$, the influencer (i.e., agent $0$) cares more about the long-term behaviour of random walks over $\mathcal{G}$, which is reflected in \eqref{prob:single}; while for large stubbornness, the influencer only considers the one-hop structure of the graph, encoded by the $g_i$'s. Such a difference is illustrated in Example \ref{ex:homo_stub}. This can also be explained from the mathematical form of the cost function $f(i;\theta,\omega)$: if we expand the right-hand term of \eqref{eq:def_f}, the weights assigned to high-order terms (i.e., $(1-\theta)^k$ for $k\geq 3$), determined by the multi-hop structures of $\G$, decreases rapidly as $\theta$ grows from $0$ to $1$. 
	\end{remark}

	\section{Results for complete and ring graphs}\label{sec:special}
	
	This section will provide accurate solution to the optimization problem \eqref{eq:op_follower} for two special graphs. The first is a compete graph with a rank-$1$ weight matrix, and the second one is a ring graph with a symmetric and circulant weight matrix. 
	
	\subsection{Rank-$1$ complete graph}\label{sec:complete}
	
	In this subsection, we assume that the weight matrix $W$ is of rank-$1$, i.e., $W=\mathbf{1c}^\top$ for some $\mathbf{c}\in\mathbb{R}_+^n$ with $\mathbf{c}^\top \mathbf{1}=1$. For this case, by calculation, the following lemma is obtained.
	
	\begin{lemma}\label{le:sp_complete}
		Consider the FJ model with $W=\mathbf{1c}^\top$. It holds 
		\begin{equation}\label{eq:sp_rank1}
			\begin{aligned}
				&\sop_0=\frac{1}{n+1}\\
				&+ \frac{\omega\left\{\sum\limits_{j\in\V}c_j\theta_j\sum\limits_{j\in\mathcal{S}}(1-\theta_j)+\sum\limits_{j\in\V}(1-\theta_j)\sum\limits_{j\in\mathcal{S}}c_j(1-\theta_j)\right\}}{(n+1)\left\{\sum\limits_{j\in \V}c_j\theta_j+\omega\sum\limits_{j\in\mathcal{S}}c_j(1-\theta_j)\right\}},
			\end{aligned}
		\end{equation}
	\end{lemma}
	
	The proof of Lemma \ref{le:sp_complete} is in Appendix \ref{ap:3}. According to Lemma \ref{le:sp_complete}, the solution to the problem \eqref{eq:op_follower} for special parameter settings can be given explicitly, as shown in the following corollary.
	
	\begin{corollary}\label{prop:rank1-special}
		Consider the problem \eqref{eq:op_follower} with $W=\mathbf{1c}^\top$. Let $\theta_{i_1},\dots, \theta_{i_n}$ and $c_{j_1},\dots, c_{j_n}$ be permutations of $\{\theta_i:i\in\V\}$ and $\{c_j:j\in\V\}$, such that $\theta_{i_1}\leq\theta_{i_2}\leq\dots<\theta_{i_n}$ and $c_{j_1}\geq c_{j_2}\geq \dots\geq c_{j_n}$. The following statements holds.
		\begin{enumerate}
			\item\label{case:same_stub} If $\theta_1=\dots=\theta_n$, $\{c_{j_1},c_{j_2},\dots,c_{j_K}\}$ is a solution to the problem \eqref{eq:op_follower};
			\item\label{case:same_centr} If $c_1=\dots=c_n$, $\{\theta_{i_1},\theta_{i_2},\dots,\theta_{i_K}\}$ is a solution to the problem \eqref{eq:op_follower};
			\item\label{case:same_prod} If $c_1(1-\theta_1)=\dots=c_n(1-\theta_n)$, $\{\theta_{i_1},\theta_{i_2},\dots,\theta_{i_K}\}$ is a solution to the problem \eqref{eq:op_follower}.
		\end{enumerate}
	\end{corollary}
	
	The proof of Corollary \ref{prop:rank1-special} is simply by checking the monotonicity of $\sop_0$ w.r.t $\sum_{j\in\mathcal{S}}(1-\theta_j)$ or $\sum_{j\in\mathcal{S}}c_j$ for the special cases, and is omitted here.
	
	\begin{remark}
		The first two cases in Corollary \ref{prop:rank1-special} are intuitive: 1) When the agents in $\V$ have same stubbornness, they are equally willing to listen to the opinion of agent $0$, and as such, agent $0$ needs to choose the agents with highest network centralities (i.e., $c_i$) as her followers to maximize her influence. 2) When the network centralities of agent in $\V$ are homogenous, agent $0$ need to choose followers who are the easiest to be influenced, i.e., the ones with lowest stubbornness.  
	\end{remark}
	
	\begin{remark}
		Case~\ref{case:same_prod}) in Corollary~\ref{prop:rank1-special} corresponds to the scenario where agents with higher network centrality are more stubborn—a pattern supported by empirical studies, as opinion leaders (typically centrally positioned) are less susceptible to peer influence \cite{van2011opinion}. Interestingly, in this case, it is more effective for agent~$0$ to link to easily influenced agents rather than those with high centrality.
	\end{remark}
	
	For generic parameters $\{\theta_i\}$ and $\{c_j\}$, the problem \eqref{eq:op_follower} is in fact a constrained hyperbolic $0-1$ programming problem \cite{hansen1991hyperbolic}. To be more clear, define
	\begin{equation}\label{eq:parameters}
		\begin{aligned}
			b_i&=[\sum_{j\in\V}c_j\theta_j+c_i\sum_{j\in\V}(1-\theta_j)](1-\theta_i),\quad i\in[n],\\
			a_0&=\sum_{j\in\V}c_j\theta_j, \quad a_i=\omega (1-\theta_i), \quad i\in[n].
		\end{aligned}
	\end{equation}
	With \eqref{eq:sp_rank1}, the problem \eqref{eq:op_follower} becomes
	\begin{equation}\label{eq:op_rank1}
		\begin{aligned}
			&\mathrm{Maximize}_{\mathcal{S}}\quad\frac{\sum_{i\in\mathcal{S}}b_i}{a_0+\sum_{i\in\mathcal{S}}a_i}\\
			&\quad\mathrm{s.t.} \quad \quad |\mathcal{S}|= K
		\end{aligned}
	\end{equation}
	Here, the constraint becomes equality due to the monotonicity of $\sop_0$ w.r.t $\mathcal{S}$. By borrowing some techniques from \cite{hansen1991hyperbolic}, we can prove that the problem \eqref{eq:op_rank1} can be solved in polynomial time, as shown in the following theorem.
	
	\begin{theorem}\label{th:complete}
		Consider the optimization problem \eqref{eq:op_rank1} with the parameters defined in \eqref{eq:parameters}. There is an algorithm to solve the problem in $O(n^3\log n)$ time.
	\end{theorem}
	
	The proof of Theorem \ref{th:complete} is given in Appendix \ref{ap:3}. Equivalently, Theorem \ref{th:complete} means that for rank-$1$ compete graphs, an accurate solution of the problem \eqref{eq:op_follower} can be found in $O(n^3\log n)$ time. Note that the solution-seeking algorithm in fact follows the procedures 1)-5) in the proof.

	\subsection{Ring graph}\label{sec:ring}
	
	
	
	In this subsection, we assume that $\G$ is a ring graph, with each node having a self-loop. The weight matrix $W$ is a symmetric circulant matrix, and has the following form for odd and even number of agents.
	\begin{itemize}
		\item[(a)] \textbf{Odd}. If $n=2s-1$ for some $s>1$, $W=C(w_0,w_1,\dots,w_{s-1},w_{s-1},\dots,w_1)$, with $w_0, w_1>0$, and $w_0+2\sum_{\ell=1}^{s-1}w_\ell=1$;
		\item[(b)] \textbf{Even}. If $n=2s$ for some $s>1$, $W=C(w_0,w_1,\dots,\allowbreak w_{s-1},w_s,w_{s-1},\dots,w_1)$, with $w_0, w_1>0$, and $w_0+w_s+2\sum_{\ell=1}^{s-1}w_\ell=1$.
	\end{itemize}
	We also assume that all agents have the same stubbornness, i.e., $\theta_i=\theta\in(0,1), \forall i\in\V$. Moreover, we consider that case that $K=2$. 
	
	Due to the symmetry of the graph $\G$ encoded in $W$, and that all the agents have same stubbornness, without loss of generality, we can let $1\in\mathcal{S}^\ast$, where $\mathcal{S}^\ast$ is a solution to the optimization problem \eqref{eq:op_follower}. The problem then boils down to finding the other agent in $\s^\ast$. Utilizing the Sherman-Morrison formula, the following lemma is obtained.
	
	\begin{lemma}\label{le:circu}
		Consider the FJ model with $\theta_i=\theta\in(0,1),\forall i\in\V$. The weight matrix $W$ is assumed to be a symmetric circlulant matrix. For this case, the inverse of the matrix $M:=I-(1-\theta)W$ exists and is circulant, denoted as $M^{-1}:=C(m_0,m_1,\dots,m_{n-1})\in\mathbb{R}^{n\times n}$. Let 
		\[j^\ast \in\mathop{\arg\min}_{j\in\V} m_{j-1},\]
		the set $\s^\ast=\{1,j^\ast\}$ is then a solution to the problem \eqref{eq:op_follower} with $K=2$.
	\end{lemma}
	
	The proof of Lemma \ref{le:circu} is given in Appendix \ref{ap:4}. Based on Lemma \ref{le:circu}, if we further assume that the link weights of $\G$ decrease with the distance between the nodes in the cycle where the nodes are located, the solution of \eqref{eq:op_follower} can be explicitly shown as follows.
	
	\begin{theorem}\label{th:cycle}
		Consider the FJ model with $\theta_i=\theta\in(0,1), \forall i\in\V$. For the optimization problem \eqref{eq:op_follower} with $K=2$, let $\mathcal{S}^\ast=\{1,j^\ast\}$ be a solution. 
		\begin{enumerate}
			\item \textbf{Odd}. If $W=C(w_0,w_1,\dots,w_{s-1},w_{s-1},\dots,w_1)$, with $w_0\geq w_1\geq  w_2\geq  \dots\geq w_{s-1}\geq 0$, $w_1>w_{s-1}$ and $w_0+2\sum_{\ell=1}^{s-1}w_\ell=1$, then $j^\ast=s$ or $s+1$;
			\item \textbf{Even}. If $W=C(w_0,w_1,\dots,w_{s-1},w_s,w_{s-1},\dots,w_1)$, with $w_0\geq w_1\geq w_2\geq  \dots\geq w_{s-1}\geq 0$, $w_1>w_{s-1}$ and $w_0+w_s+2\sum_{\ell=1}^{s-1}w_\ell=1$, then $j^\ast=s+1$.
		\end{enumerate}
	\end{theorem}
	
	The proof of Theorem \ref{th:cycle} is given in Appendix \ref{ap:4}.
	
	Theorem \ref{th:cycle} means that for the symmetric ring graph $\G$, the optimal strategy for agent $0$ to maximize her social power is to add links to two agents that have the largest distance from each other on the graph. 
	
	\begin{remark}
		The monotonicity requirement of the link weights is necessary for Theorem \ref{th:cycle}. This can be seen by considering the graph with $\V=[12]$ and $W=C(0.16,0.14,0.28,\mathbf{0}_7^\top,0.28,0.14)$, for which the monotonicity is not satisfied. Let $\theta=0.1$ and $\omega=0.2$. Then, the solution of \eqref{eq:op_follower} with $K=2$ is $\{1,6\}$, instead of $\{1,7\}$ implied by Theorem $\ref{th:cycle}$. 
	\end{remark}
	
	\begin{remark}\label{re:ring_multi}
		From Theorem \ref{th:cycle}, a conjecture could be drawn for the case of $K\geq 2$: to maximize the social power, agent $0$ should choose agents that are sufficiently dispersed over the cycle. This conjecture is supported by Example \ref{ex:ring_multi}. Note that in Example \ref{ex:ring_multi}, the monotonicity assumption in Theorem \ref{th:cycle} is no longer required, which indicates that choosing agents that are maximally dispersed over the cycle of a symmetric ring graph (with homogenous stubbornness of agents) might be at least a near-optimal strategy for agent $0$.    
	\end{remark}
	
	
	\section{Numerical Examples}\label{sec:example}
	
	The first example is to show the efficiency of the greedy algorithm, i.e., Algorithm \ref{alg:1}.
	
	\begin{figure}[t]
		\centering
		\includegraphics[trim=0cm 0cm 0cm 0cm,clip=true,width=8cm]{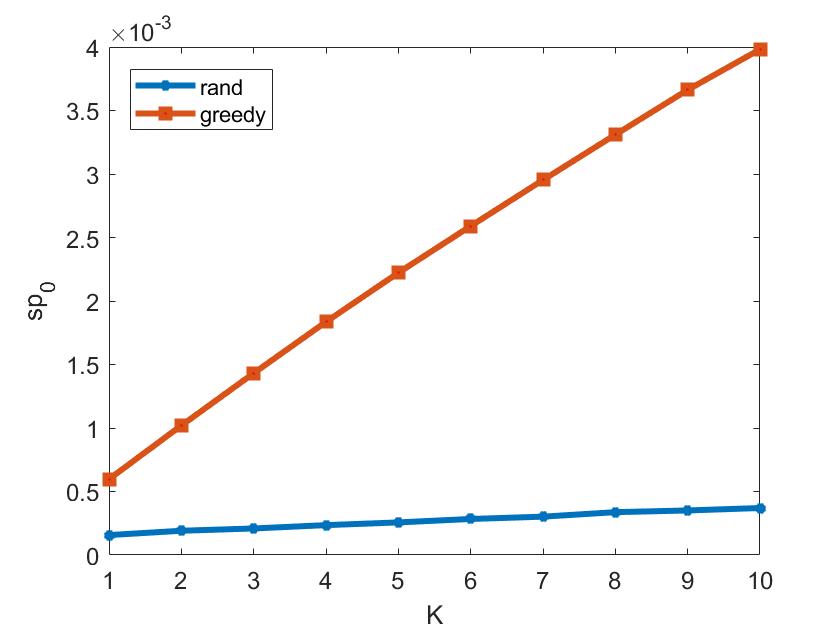}
		\caption{Simulation for Example \ref{ex:gd}. Blue and red curve corresponds to random selection of $\s$ and Algorithm \ref{alg:1}, respectively.}
		\label{fig:gd}
	\end{figure}
	
	\begin{example}\label{ex:gd}
		Consider the social power optimization problem \eqref{eq:op_follower} on the LastFM Asia Social Network \cite{feather}, in which $\V=7624$ and $\E=27806$. The weight matrix $W$ is obtained by normalize the adjacency matrix of the network. For each $i\in\V$, $\theta_i$ is randomly chosen from $[0.1,1]$. Let $\omega=0.2$. Such a large network size makes it impractical to implement an exhaustive search for the accurate solution of \eqref{eq:op_follower}. For each $K=1,2,\dots, 10$, we run two algorithms to select $\s$ and compute the social power of agent $0$ by using Algorithm \ref{alg:1} or randomly select $K$ agents from $\V$ to consist $\s$, and repeat this process for $100$ times to obtain an average value of $\sop_0$. The simulation results is shown in Fig. \ref{fig:gd}. It can be seen that for each $K$, the optimized social power generated by Algorithm~\ref{alg:1} is obviously higher than that of the random selection. Note that the growth of both the optimized $\sop_0$ (red curve) and the randomized one (blue curve) w.r.t $K$ seem to be close to (but not exact) a linear growth. This might be because the maximum number of $K$ in the simulation, i.e., $10$, is rather small compared to the network size, and the marginal increase of social power brought by the $10$ selected agents are close to each other for this special network structure.
	\end{example}
	
	\begin{figure}[t]
		\centering
		\includegraphics[trim=0cm 0cm 0cm 0cm,clip=true,width=8cm]{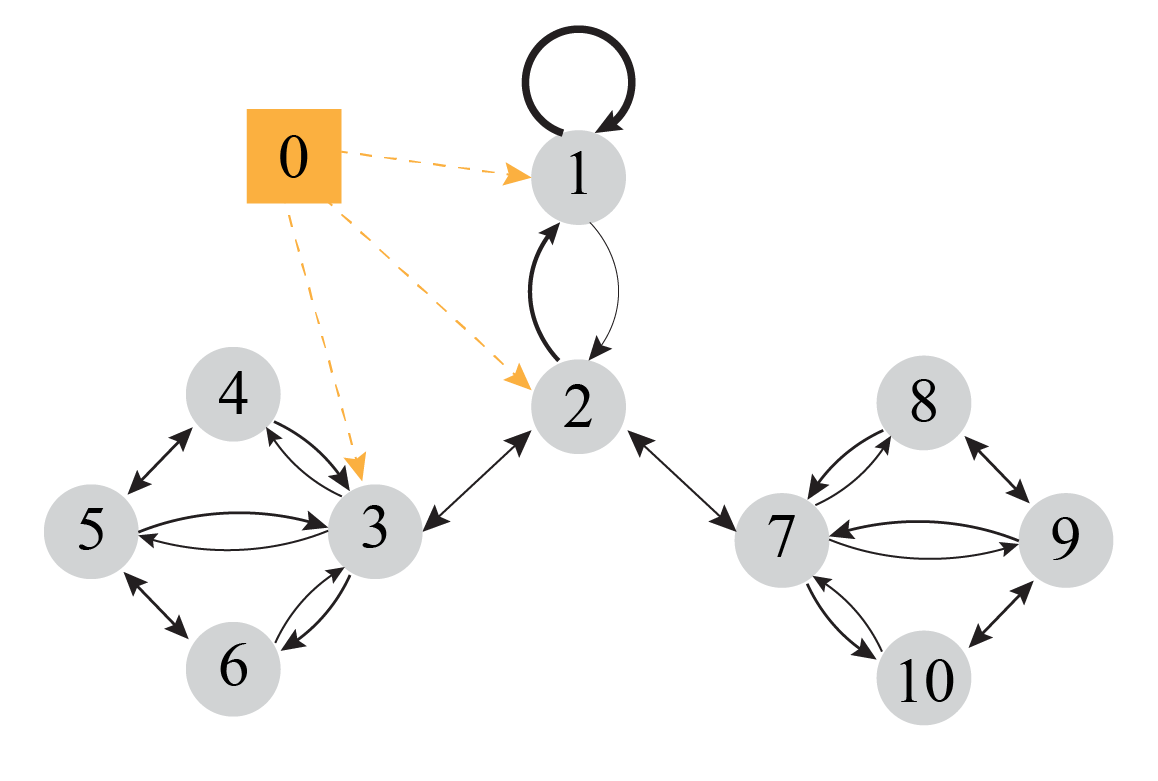}
		\caption{Graph for Example \ref{ex:homo_stub}. Edge thickness represent the weight value.}
		\label{fig:graph_homo_stub}
	\end{figure}
	
	\begin{figure}[t]
		\centering
		\includegraphics[trim=0cm 0cm 0cm 0cm,clip=true,width=8cm]{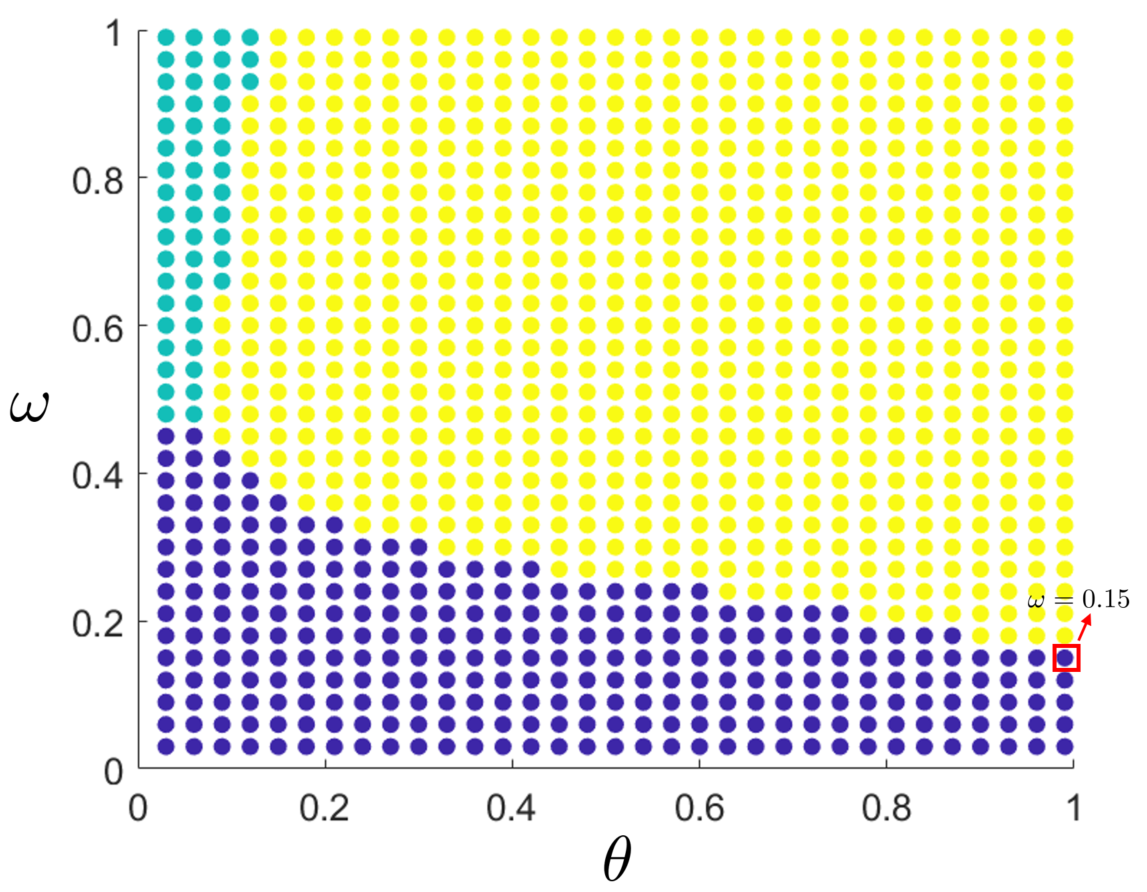}
		\caption{Results for Example \ref{ex:homo_stub}: solution of \eqref{eq:op_follower} for different $\theta$ and $\omega$. The solutions are represented by different colors, with blue color for agent $1$, cyan color for agent $2$, and yellow color for agent $3$ (and also agent $7$, as they are solutions simultaneously).}
		\label{fig:homo_stub}
	\end{figure}
	
	We present the following example to illustrate the lemmas in Section \ref{subsec:homo_stub}.
	
	\begin{example}\label{ex:homo_stub}
		Consider the social power optimization problem \eqref{eq:op_follower} with $n=10$ and $K=1$. The stubbornness of each agent is the same, denoted as $\theta$. The graph $\G$ is shown in Fig. \ref{fig:graph_homo_stub}, with the weight matrix $W$ defined as: $W_{11}=0.89, W_{12}=0.11, W_{21}=0.5, W_{23}=W_{27}=0.25$, and for each of the other nodes, the weights of all outgoing links are the same. The stubbornness $\theta$ and link weight of the influencer $\omega$ both take values from $\{0.03,0.06,\dots, 0.99\}$. For each pair of $(\theta,\omega)$, the solution of \eqref{eq:op_follower} is obtained by exhaustive search. The results are shown in \ref{fig:homo_stub}. It can be seen that
		\begin{itemize}
			\item when the stubbornness is close to $0$, say, $\theta\leq 0.06$, the optimal solution changes from $1$ to $3$ (or $7$) as $\omega$ grows. This is predicted by Remark \ref{re:small_stub}, since for small $\omega$ the agent with small expected returning time (i.e., $\mathbb{E}[\tau_{ii}]$), which is agent $1$ for this graph, should be chosen; and for large $\omega$, the selected agent should be the one hit in minimum expected time on average  by random walks from the other agents, which is agent $2$. 
			\item as $\theta$ grows, the solution becomes agent $1$ for small $\omega$ and $3$ for large $\omega$. In particular, for $\theta=0.99$, the solution changes when $\omega$ changes from $0.15$ to $0.18$. Note that for $\omega=0.15$, it holds $g_3=1.25<1.2565=g_3$, and for $\omega=0.18$, it holds $g_1=1.2298<1.25=g_3$. Therefore, Proposition \ref{prop:big_stub} is validated.
		\end{itemize}
	\end{example}
	
	The last example provides numerical evidence to support Theorem \ref{th:cycle} and Remark \ref{re:ring_multi}.
	
	\begin{figure}[t]
		\centering
		\includegraphics[trim=0cm 0cm 0cm 0cm,clip=true,width=8cm]{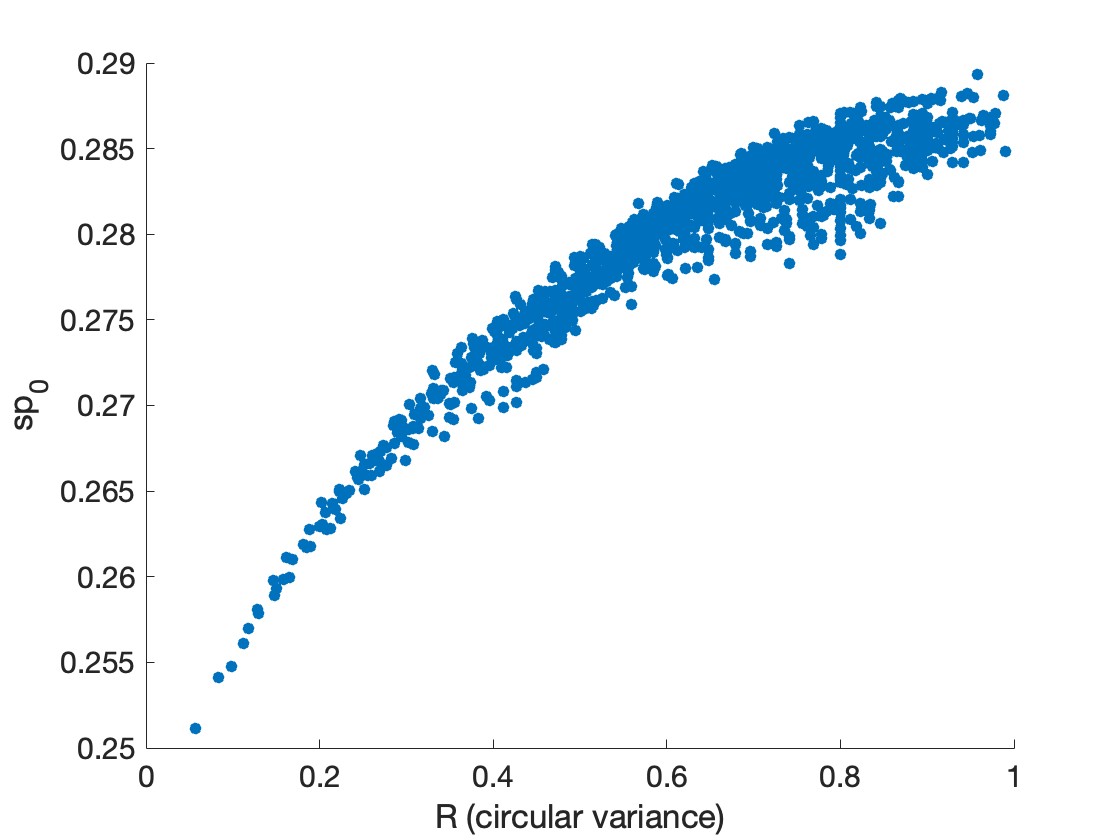}
		\caption{Results for Example \ref{ex:ring_multi}. Each scatter plot point corresponds to a subset $\mathcal{S}$ chosen by agent $0$ (different subsets with the same $R$ and $\sop_0$ are merged into one point).}
		\label{fig:ring_multi}
	\end{figure}
	
	\begin{example}\label{ex:ring_multi}
		Consider the social power optimization problem \eqref{eq:op_follower} with $n=26$ and $K=5$. The stubbornness of all agents are $0.1$. The weights of links added by agent $0$ are $\omega=0.2$. The graph $\G$ is a ring, and the weight matrix is $C(0.02,\mathbf{w}^\top,\mathbf{0}_{17}^\top,\mathbf{w}^\top)$, with $\mathbf{w}^\top=[0.17,0.11,0.09,0.12]$. Each agent $i$ is associated with a coordinate $(\cos(\phi_i), \sin(\phi_i))$ on the unit circle, where $\phi_i = \frac{2\pi(i-1)}{26}$, reflecting its position on the cycle graph. For any subset $\mathcal{S} \subseteq \{1,\dots,n\}$ with $|\mathcal{S}| = 5$, let the circular variance be defined as
		\[
		R(\mathcal{S}) = 1 - \frac{1}{|\mathcal{S}|} \sqrt{\left( \sum_{i \in \mathcal{S}} \cos(\phi_i) \right)^2 + \left( \sum_{i \in \mathcal{S}} \sin(\phi_i) \right)^2}.
		\]
		This quantity measures how evenly the agents in $\mathcal{S}$ are distributed around the cycle. We evaluate all possible subsets $\mathcal{S}$ of size 5, compute their circular variances, and determine the corresponding social power $\sop_0(\mathcal{S})$ of agent 0 when selecting $\mathcal{S}$. The results, shown in Fig.~\ref{fig:ring_multi}, indicate a positive correlation between $R(\mathcal{S})$ and $\sop_0(\mathcal{S})$: the more dispersed the selected agents are, the greater the social power of agent 0.
		Similar trends are observed for other symmetric circulant weight matrices $W$. This suggests that agent 0 benefits from selecting followers that are spatially well-dispersed over the cycle---i.e., local clustering should be avoided. This observation aligns with Theorem~\ref{th:cycle} and Remark~\ref{re:ring_multi}.
	\end{example}
	
	\section{Conclusion}
	
	This paper examined the problem of maximizing social power for an external influencer, modeled as a fully stubborn agent in the FJ model. The influencer could form a limited number of links with regular agents, subject to a budget constraint. To address this problem, a greedy algorithm was developed by leveraging the monotonicity and submodularity of the payoff function. An augmented Markov chain was introduced to provide analytical insight and was further utilized to save some computational burden for large networks. Several specific network topologies were analyzed. For complete graphs with a rank-one weight matrix, it was shown that an exact solution can be obtained in polynomial time. For symmetric ring graphs with circulant weight matrices and homogeneous agent stubbornness, it was demonstrated that when at most two agents are selected, the optimal choice is to connect to agents that are sufficiently dispersed across the network. While numerical examples suggested this pattern may extend to larger selections, a formal theoretical proof remained an open direction for future work.

	\appendix
	\section{Proofs for Subsection \ref{sec:alg}}\label{ap:1}
	
	\noindent\textbf{Proof of Proposition \ref{prop:monotone}}. In \eqref{eq:sol_FJ},  it is easy to see that $H(\s)$ is strictly sub-stochastic. Moreover, it holds $\mathbf{0}\leq H(\s\cup\{i\})\leq H(\s)$. Therefore, it holds
	\[P(\s)=\sum_{k=0}^\infty H(\s)^k\Theta\geq \sum_{k=0}^\infty H(\s\cup \{i\})^k\Theta=P(\s\cup \{i\}). \]
	Due to the fact that $\bar{P}(\s)$ and $\bar{P}(\s\cup \{i\})$ are both stochastic, it must be $\mathbf{p}_0(\s)\leq \mathbf{p}_0(\s\cup \{i\})$. Notice that $P(\s)\not= P(\s\cup \{i\})$, the inequality must hold strictly, which gives the desired conclusion.\hfill$\square$ 
	\hfill\break
	
	Before going to the proof of Proposition \ref{prop:submodular}, some notations need to be given. Consider a square matrix $A\in\mathbb{R}^{|\V|\times|\V|}$. Let $\mathcal{P}=i_0\to i_1\to \dots \to i_\ell$ be a path on the graph $\G$. Denote $v_k(\mathcal{P})=i_k$ as the $(k+1)$-th node that $\mathcal{P}$ goes through, and specially, $v_0(\mathcal{P})=i_0$ as the starting node of $\mathcal{P}$. Denote $A_{\mathcal{P}}:=A_{i_\ell i_{\ell-1}}\dots A_{i_2 i_1}A_{i_1i_0}$ as the product of the entries corresponding to all edges (counted with multiplicity) on the path $\mathcal{P}$. For any $i\in\V$, denote $A_{\mathcal{P}\cap i}=\Pi_{\substack{k\in [\ell]\\ i_k=i} }A_{i_ki_{k-1}}$ as the product of the entries corresponding to all edges (counted with multiplicity) on the path $\mathcal{P}$ that start from $i$, and denote $A_{\mathcal{P}\backslash i}=A_\mathcal{P}/A_{\mathcal{P}\cap i}$ as the entry product corresponding to all edges on the path $\mathcal{P}$ with its starting point not being $i$. 
	\hfill\break
	
	\noindent\textbf{Proof of Proposition \ref{prop:submodular}}. Consider two arbitrary set $\s, \mathcal{T}$ such that $\mathcal{T}\subset\s$. We need to prove that for any $i\in \V\backslash\s$, it holds 
	\[\sop_0(\s\cup\{i\})-\sop_0(\s)\leq \sop_0(\T\cup \{i\})-\sop_0(\T). \]
	Due to the stochasticity of $\bar{P}(\s), \bar{P}(\s\cup \{i\}), \bar{P}(\T)$ and $\bar{P}(\T\cup \{i\})$, it suffices to prove that for any $j,h\in\V$, it holds
	\begin{equation}
		[P(\s)]_{jh}-[P(\s\cup \{i\})]_{jh}\leq [P(\T)]_{jh}-[P(\T\cup \{i\})]_{jh}.
	\end{equation}
	Observe that
	\begin{equation}\label{eq:margi_es}
		\begin{aligned}
			&\sum_{k=0}^\infty ([H(\s)^k]_{jh}-[H(\s\cup \{i\})^k]_{jh})\\
			&=\sum_{k=0}^\infty\sum_{\substack{|\mathcal{P}|=k\\ v_0(\mathcal{P})=h, v_k(\mathcal{P})=j}} \left([H(\s)]_\mathcal{P}-[H(\s\cup \{i\})]_\mathcal{P}\right)\\
			&= \sum_{k=0}^\infty\sum_{\substack{|\mathcal{P}|=k\\ v_0(\mathcal{P})=h\\ v_k(\mathcal{P})=j}}\left([H(\s)]_{\mathcal{P}\cap i}-[H(\s\cup \{i\})]_{\mathcal{P}\cap i}\right)[H(\s)]_{\mathcal{P}\backslash i}\\
			&= \sum_{k=0}^\infty\sum_{\substack{|\mathcal{P}|=k\\ v_0(\mathcal{P})=h\\ v_k(\mathcal{P})=j}}\left([H(\T)]_{\mathcal{P}\cap i}-[H(\T\cup \{i\})]_{\mathcal{P}\cap i}\right)[H(\s)]_{\mathcal{P}\backslash i}\\
			&\leq \sum_{k=0}^\infty\sum_{\substack{|\mathcal{P}|=k\\ v_0(\mathcal{P})=h\\ v_k(\mathcal{P})=j}}\left([H(\T)]_{\mathcal{P}\cap i}-[H(\T\cup \{i\})]_{\mathcal{P}\cap i}\right)[H(\T)]_{\mathcal{P}\backslash i}\\
			&= \sum_{k=0}^\infty ([H(\T)^k]_{jh}-[H(\T\cup \{i\})^k]_{jh}).
		\end{aligned}
	\end{equation}
	Therefore, we have 
	\begin{equation*}
		\begin{aligned}
			&[P(\s)]_{jh}-[P(\s\cup \{i\})]_{jh}\\
			&= \sum_{k=0}^\infty ([H(\s)^k]_{jh}-[H(\s\cup \{i\})^k]_{jh})\theta_h\\
			&\leq [P(\T)]_{jh}-[P(\T\cup \{i\})]_{jh}.
		\end{aligned}
	\end{equation*}
	The proof is then completed.\hfill$\square$
	\hfill\break
	
	
	To prove Theorem \ref{th:high_probability}, the Chernoff bound needs to be introduced at first.
	
	\begin{lemma}[Chernoff bound]\label{le:chernoff}
		Suppose that $X_1$, $\dots$, $X_n$ are independent Bernoulli random variables such that $\mathbb{P}\{X_i = 1\} = p_i = 1 - \mathbb{P}\{X_i=0\}$. Let $X := \sum_{i=1}^n X_i$ and $\mu := \mathbb{E}\{X\} = \sum_{i=1}^n p_i$. Then for $0< \epsilon < 1$,
		\begin{align*}
			\mathbb{P}\{|X - \mu| \ge \epsilon \mu\} \le 2 e^{-\mu\epsilon^2/3},
		\end{align*}
	\end{lemma}
	
	\noindent\textbf{Proof of Theorem \ref{th:high_probability}}. Note that $\sop_0=\sop_0^\ell (\s)+\sop_0^{\ell c}(\s)$, with
	\begin{equation}\label{eq:sp_truncate}
		\begin{aligned}
			\sop_0^{\ell c}(\s)&:= \frac{1}{n}\mathbf{1}^\top \sum_{k=\ell+1}^\infty H(\s)^k(I-\Theta)\omega\sum_{j\in\s}\mathbf{e}_j.
		\end{aligned}
	\end{equation}
	Consider a finite-length random walk $\mathcal{Y}=\{Y_t: t=0,1,\dots, \ell\}$ generated by the Markov chain with the transition matrix as $T$ (defined by \eqref{eq:trans_mat}), and the starting state is selected uniformly from $\V$. Note that $\sop_0^\ell(\s)$ is in fact the probability that $\mathcal{Y}$ is absorbed by the state $0$, i.e., 
	\[\sop_0^\ell(\s)=\mathbb{P}(\exists t\leq \ell \text{ s.t. } Y_t=0)=\mathbb{E}[\mathbf{1}\{\exists t\leq \ell \text{ s.t. } Y_t=0\}]. \]
	Therefore, due to the fact the random walks $\mathcal{Y}^{(k)}$ in Algorithm \ref{alg:2} are $i.i.d$, we can obtained from the Lemma \ref{le:chernoff}
	that
	\begin{equation*}\label{eq:hpd_1}
		\mathbb{P}(|r \sop_0^\text{apx}(\s)-r \sop_0^\ell(\s)|\geq \epsilon r\sop_0^\ell(\s) )\leq 2 e^{-r  \sop_0^\ell(\s) \epsilon^2/3}.
	\end{equation*}
	Hence, for $r \geq 3 \log (2/\delta) / ( \sigma^2 \epsilon^2 \sop_0^\ell(\s))$,
	\begin{equation*}\label{eq:hpd_1}
		\mathbb{P}(|\sop_0^\text{apx}(\s)- \sop_0^\ell(\s)|\geq \sigma \epsilon \sop_0^\ell(\s) )\leq 2 e^{-r  \sop_0^\ell(\s) \sigma^2 \epsilon^2/3} \le \delta.
	\end{equation*}
	On the other hand, for $\ell \geq  [\log(\theta(1-\sigma)\epsilon \sop_0^\ell(\s))-\log \omega]/[\log(1-\theta)]-2$,
	\begin{equation}\label{eq:hpd_2}
		\begin{aligned}
			&\sop_0^{\ell c}(\s)\leq \frac{1}{n}\|\mathbf{1}^\top\|_{\infty}\sum_{k=\ell+1}^\infty \|H(\s)\|_\infty^k\|I-\Theta\|_\infty \omega\|\sum_{j\in\s}\mathbf{e}_j\|_\infty\\
			&\leq \omega\sum_{k=\ell+1}^\infty (1-\theta)^{k+1}=\frac{\omega(1-\theta)^{\ell+2}}{\theta}\leq (1-\sigma)\epsilon \sop_0^\ell(\s),
		\end{aligned}
	\end{equation}
	where the first inequality is from 
	\[\|H(\s)\|_\infty\leq \|I-\omega\sum_{j\in\s}\mathbf{e}_j\mathbf{e}_j^\top\|_\infty \|I-\Theta\|_\infty \|W\|_{\infty}\leq 1-\theta. \]
	Therefore,
	\begin{equation*}
		\begin{aligned}
			&\mathbb{P}(|\sop_0^\text{apx}(\s)-\sop_0(\s)|< \epsilon \sop_0^\ell(\s))\geq 1 \\
			& -\mathbb{P}(|\sop_0^\text{apx}(\s)-\sop_0^\ell(\s)|+|\sop_0^\ell(\s)-\sop_0(\s)|\geq \epsilon \sop_0^\ell(\s))\\
			&\geq 1-\mathbb{P}(|\sop_0^\text{apx}(\s)-\sop_0^\ell(\s)|\geq \sigma \epsilon \sop_0^\ell(\s))\geq 1-\delta.
		\end{aligned}
	\end{equation*}
	The proof is then completed.

	\section{Proofs for Subsection \ref{subsec:homo_stub}}\label{ap:equiv_markov}
	
	\noindent\textbf{Proof of Proposition \ref{prop:equiv_single}}. If suffices to prove that for $\forall i\in\V$,
	\begin{equation}\label{eq:f_0}
		f(i;0,\omega)=\frac{1}{n}\sum_{j\not= i}\mathbb{E}_{\mathcal{Y}(\G, \mathbf{0})}[\tau_{ji}]+\frac{1-\omega}{\omega}\mathbb{E}_{\mathcal{Y}(\G, \mathbf{0})}[\tau_{ii}]+1.
	\end{equation}
	We simply use $\mathcal{Y}, \bar{\mathcal{Y}}$ to represent the augmented Markov chains $\mathcal{Y}(\mathcal{G},\mathbf{0})$ and $\bar{\mathcal{Y}}(\bar{\G},\mathbf{0})$. Consider $\bar{\mathcal{Y}}=\{\bar{Y}_t:t\in\mathbb{Z}_{\geq 0}\}$ with $\s=\{i\}$. Note that with probability $1$, $\bar{\mathcal{Y}}$ will be  absorbed by the state $i'$. For each $j\in\V$, define
	\[\bar{\tau}_j:=\min\{t:\bar{Y}_t=i', \bar{Y}_0=j\}. \]
	According to the proof of Lemma \ref{le:equiv_markov}, 
	\begin{equation}\label{eq:exp_sum}
		f(i;0,\omega)=\frac{1}{n}\sum_{j\in\V}\mathbb{E}_{\bar{\mathcal{Y}}}[\bar{\tau}_{j}].
	\end{equation}
	We now consider $\bar{\mathcal{Y}}$ with $\bar{Y}_0=j$. Let $N$ be the number of times that $\bar{\mathcal{Y}}$ hits $i$ before absorbed by $i'$, and for $k\in [N]$ define
	\[t_k:=\min\{t-t_{k-1}: t>t_{k-1}, \bar{Y}_t=i\} \]
	as time interval between the $(k-1)$th and $k$th hitting at $i$, where $t_0:=0$. We then have $\tau_j=\sum_{k=1}^Nt_k+1$
	By the law of total expectation,
	\begin{equation}
		\begin{aligned}
			\mathbb{E}[\bar{\tau}_j]&=\mathbb{E}_N[\mathbb{E}[\bar{\tau}_j|N]]\\
			&=\sum_{k=1}^\infty \mathbb{P}(N=k)\left(\sum_{\ell=1}^k \mathbb{E}[t_{\ell}|N=k]+1\right). 
		\end{aligned}
	\end{equation}
	Note that $\bar{\mathcal{Y}}$ is an homogenous Markov chain, and under the condition $N=k$, for any $\ell_1,\ell_2\in\{2,\dots,N\}$, the distribution of $t_{\ell_1}$ is the same as that of $t_{\ell_2}$. In fact, for any $\ell\in\{2,\dots,N\}$, the distribution of $t_\ell$ on condition that $N=k\geq 2$ is the same as that of $\tau_{ii}$, which is defined over the Markov chain $\mathcal{Y}$. Moreover, if $j\not=i$, the distribution of $t_1$ is the same as that of $\tau_{ji}$; otherwise if $j=i$, $t_1=0$. Therefore, for $j\not=i$, 
	\begin{equation}\label{eq:sum_1}
		\begin{aligned}
			\mathbb{E}[\bar{\tau}_j]&=\sum_{k=1}^\infty \mathbb{P}(N=k)(\mathbb{E}[\tau_{ji}]+(k-1)\mathbb{E}[\tau_{ii}]+1)\\
			&=\mathbb{E}[\tau_{ji}]+\sum_{k=1}^\infty (k-1)\mathbb{P}(N=k)\mathbb{E}[\tau_{ii}]+1\\
			&=\mathbb{E}[\tau_{ji}]+\sum_{k=1}^\infty (k-1)(1-\omega)^{k-1}w\mathbb{E}[\tau_{ii}]+1\\
			&=\mathbb{E}[\tau_{ji}]+\frac{1-\omega}{\omega}\mathbb{E}[\tau_{ii}]+1,
		\end{aligned}
	\end{equation}
	where the third equality is because
	\begin{equation*}
		\begin{aligned}
			\mathbb{P}(N=k)&=\mathbb{P}(\bar{Y}_{t_k+1}=i')\Pi_{\ell=1}^{k-1}\mathbb{P}(\bar{Y}_{t_\ell+1}\not=i')\\
			&=\omega(1-\omega)^{k-1}.
		\end{aligned}
	\end{equation*}
	On the other hand, if $j=i$, it holds
	\begin{equation}\label{eq:sum_2}
		\begin{aligned}
			\mathbb{E}[\bar{\tau}_i]&=\sum_{k=1}^\infty \mathbb{P}(N=k)((k-1)\mathbb{E}[\tau_{ii}]+1)\\
			&= \frac{1-\omega}{\omega}\mathbb{E}[\tau_{ii}]+1.
		\end{aligned}
	\end{equation}
	Combining \eqref{eq:exp_sum} with \eqref{eq:sum_1}-\eqref{eq:sum_2}, the equation \eqref{eq:f_0} is obtained. Then from Lemma \ref{le:equiv_markov}, the desired conclusion is reached.
	\hfill$\square$
	\\ \hfill\\
	
	\noindent\textbf{Proof of Proposition \ref{prop:big_stub}}. Denote $H_i:=(I-\omega \mathbf{e}_i\mathbf{e}_i^\top)W$. It holds
	\begin{equation}
		\begin{aligned}
			&f(i;\theta,\omega)=\frac{1}{n}\sum_{k=0}^\infty (1-\theta)^k \mathbf{1}^\top H_i^k\mathbf{1} \\
			&\quad= 1+(1-\theta)(1-\frac{\omega}{n})+\frac{1}{n}\sum_{k=2}^\infty (1-\theta)^k \mathbf{1}^\top H_i^k\mathbf{1},
		\end{aligned}
	\end{equation}
	where the second equality is due to the fact that $W\mathbf{1}=\mathbf{1}$. From the equivalence between the problems $\eqref{prob:markov}$ and $\eqref{prob:homo_stub}$, the problem becomes 
	\begin{equation}\label{prob:big_stub}
		\mathrm{Minimize}_{i\in\V} \quad   \mathbf{1}^\top H_i^2\mathbf{1}+\sum_{k=1}^\infty (1-\theta)^k \mathbf{1}^\top H_i^{k+2}\mathbf{1}. 
	\end{equation}
	By calculation,
	\begin{equation*}
		\begin{aligned}
			\mathbf{1}^\top H_i^2\mathbf{1}&=\mathbf{1}^\top W\mathbf{1}-\omega\mathbf{1}^\top W\mathbf{e}_i-\omega\mathbf{e}_i^\top W\mathbf{1}+\omega^2\mathbf{e}_i^\top W\mathbf{e}_i \\
			&=n-\omega-\omega\sum_{j\in\V}W_{ji}+\omega^2 W_{ii}\\
			&=n-\omega-\omega g_i.
		\end{aligned}
	\end{equation*}
	Therefore, the optimization problem \eqref{prob:big_stub} is equivalent to
	\begin{equation}\label{prob:big_stub2}
		\mathrm{Minimize}_{i\in\V} \quad   -g_i+ \sum_{k=1}^\infty (1-\theta)^k \mathbf{1}^\top H_i^{k+2}\mathbf{1}.
	\end{equation}
	The problem to \eqref{prob:big_stub2} is $i^\ast$. This is because for any $i\not= i^\ast$, it holds
	\begin{equation*}
		\begin{aligned}
			&g_{i^\ast}-g_i+ \sum_{k=1}^\infty (1-\theta)^k \mathbf{1}^\top H_i^{k+2}\mathbf{1}- \sum_{k=1}^\infty (1-\theta)^k \mathbf{1}^\top H_{i^\ast}^{k+2}\mathbf{1}\\
			&> g_{i^\ast}-g_i- \sum_{k=1}^\infty (1-\theta)^k \mathbf{1}^\top H_{i^\ast}^{k+2}\mathbf{1}\\
			&\geq g_{i^\ast}-g_i-\|\mathbf{1}^\top\|_\infty\sum_{k=1}^\infty (1-\theta)^k\|H_i\|_{\infty}^{k+2}\|\mathbf{1}\|_\infty\\
			&\geq \delta_g-\frac{n(1-\theta)}{\theta}>0.
		\end{aligned}
	\end{equation*}
	The proof is then completed.\hfill$\square$

	\section{Proofs of Subsection \ref{sec:complete}}\label{ap:3}
	
	\noindent\textbf{Proof of Lemma \ref{le:sp_complete}}. For $W=\mathbf{1c}^\top$, we first have
	\[H(\mathcal{S})=(I-\omega\sum_{j\in\mathcal{S}}\mathbf{e}_j\mathbf{e}_j^\top)(I-\Theta)\mathbf{1c}^\top. \]
	Then,
	\begin{equation}
		\begin{aligned}
			\mathbf{p}_0&=(I-H(\mathcal{S}))^{-1}(I-\Theta)\omega\sum_{j\in\mathcal{S}}\mathbf{e}_j\\
			&= \frac{\omega\sum\limits_{j\in\mathcal{S}}c_j(1-\theta_j)\cdot (I-\omega\sum\limits_{j\in\mathcal{S}}\mathbf{e}_j\mathbf{e}_j^\top)(I-\Theta)\mathbf{1}}{1-\sum\limits_{j\in \V}c_j(1-\theta_j)+\omega\sum\limits_{j\in\mathcal{S}}c_j(1-\theta_j)}\\
			&+(I-\Theta)\omega\sum_{j\in\mathcal{S}}\mathbf{e}_j,
		\end{aligned}
	\end{equation}
	where the second equality is by applying the Sherman-Morrison formula. Accordingly, the social power of agent $0$ is 
	\begin{equation}
		\begin{aligned}
			\sop_0&=\frac{1}{n+1}+\frac{\omega}{n+1}\left\{\sum_{j\in\mathcal{S}}(1-\theta_j)\right.\\
			&+\left. \frac{\sum\limits_{j\in\mathcal{S}}c_j(1-\theta_j)[\sum\limits_{j\in\V}(1-\theta_j)-\omega\sum\limits_{j\in\mathcal{S}}(1-\theta_j)]}{\sum\limits_{j\in \V}c_j\theta_j+\omega\sum\limits_{j\in\mathcal{S}}c_j(1-\theta_j)}\right\},\\
		\end{aligned}
	\end{equation}
	which is exactly \eqref{eq:sp_rank1}, and the equality is due to $\sum_{j\in\V}c_j=1$.\hfill$\square$
	\\ \hfill \\
	
	\noindent\textbf{Proof of Theorem \ref{th:complete}}. Note that $a_0+\sum_{i\in\mathcal{S}}a_i>0$ and $\sum_{i\in\mathcal{S}}b_i>0$ hold for all $\mathcal{S}\subset \V$. Let $t^\ast$ be the optimum value of the payoff function in \eqref{eq:op_rank1}. Obviously $t^\ast>0$. Let $f(\mathcal{S}):=\sum_{i\in\mathcal{S}}b_i-t^\ast(a_0+\sum_{i\in\mathcal{S}}a_i)$. Construct the following optimization problem
	\begin{equation}\label{eq:op_substitute}
		\begin{aligned}
			&\mathrm{Maximize}_{\mathcal{S}}\; f(\mathcal{S}),\quad \mathrm{s.t.}\quad |\mathcal{S}|= K
		\end{aligned}
	\end{equation}
	For any $\mathcal{S}\subset\V$ with $|\mathcal{S}|=K$, we have the following observations: $f(\mathcal{S})\leq 0$, otherwise $t^\ast$ can not be the optimum value of \eqref{eq:op_rank1}; if $f(\mathcal{S})=0$, $\mathcal{S}$ is a solution of \eqref{eq:op_rank1}; if $f(\mathcal{S})<0$, $\mathcal{S}$ is not a solution of \eqref{eq:op_rank1}. Therefore, $\mathcal{S}$ solves \eqref{eq:op_substitute} if and only if it solves \eqref{eq:op_rank1}. The problem becomes to find an algorithm to solve \eqref{eq:op_substitute} in $O(n^3\log n)$ time.
	
	Consider the problem \eqref{eq:op_substitute}. Given $t^\ast$, we first sort the sequence of ${b_i-t^\ast a_i}$ in a decreasing order. It is easy to see that a solution to \eqref{eq:op_substitute} is to choose the agents corresponding to the first $K$ values in the sequence. Note that as $t$ goes from $0$ to $\infty$, there can be a maximum $\binom{n}{2}$ possible permutations of $\{i:i\in\V\}$ along a decreasing order of ${b_i-t a_i}$: for any $i,j\in\V$, they swap their positions in the permutated sequence when $b_i-ta_i=b_j-ta_j$, which occurs at most once. Let $t_1<t_2<\dots<t_M$ be all the swapping times, with $t_1\geq 0$ and $M\leq \binom{n}{2}$. Choose $t_1^\prime=t_1-1, t_{M+1}^\prime=t_M+1$ and $t_m^\prime=\frac{t_{m-1}+t_m}{2}, m=2,\dots, M$. We go through all $m\in[M+1]$ and execute the following steps:
	\begin{enumerate}
		\item sort $\{b_i-t_m^\prime a_i\}$ in a decreasing order: $b_{i_1}-t_m^\prime a_{i_1}\geq \dots \geq b_{i_n}-t_m^\prime a_{i_n}$;
		\item construct $\mathcal{S}_m$: $\mathcal{S}_m\leftarrow \{i_1,\dots, i_K\}$;
		\item compute: $t_m^\ast \leftarrow \frac{\sum_{i\in\mathcal{S}_m}b_i}{a_0+\sum_{i\in\mathcal{S}_m}a_i}$;
		\item sort $\{b_i-t_m^\ast a_i\}$ in a decreasing order: $b_{j_1}-t_m^\ast a_{j_1}\geq \dots \geq b_{j_n}-t_m^\ast a_{j_n}$, and construct: $\mathcal{S}_m^\ast\leftarrow \{j_1,\dots, j_K\}$;
		\item if $t_m^\ast = \frac{\sum_{i\in\mathcal{S}_m^\ast}b_i}{a_0+\sum_{i\in\mathcal{S}_m^\ast}a_i}$, return $\mathcal{S}_m^\ast$ and terminate the process; otherwise, $m\leftarrow m+1$.
	\end{enumerate}
	Due to the fact that the problem \eqref{eq:op_rank1} must have a solution, the algorithm must return a $\mathcal{S}_m^\ast$, which is exactly the solution. Note that the complexity of each loop is determined by the sorting algorithm, which can be $O(n\log n)$ \cite{knuth1998art}. As $M\leq \binom{n}{2}$, the solution $\mathcal{S}_m^\ast$ is returned in $O(n^3\log n)$ time. \hfill$\square$

	\section{Proofs for Subsection \ref{sec:ring}}\label{ap:4}
	
	\noindent\textbf{Proof of Lemma \ref{le:circu}}. For any $\mathcal{S}=\{1,j\}$, it holds
	\[\sop_0(\mathcal{S})=\frac{(1-\theta)\omega}{n+1}\mathbf{1}^\top H(\s)^{-1}(\mathbf{e}_1+\mathbf{e}_j)+\frac{1}{n+1},\]
	with 
	\[H(\s)=I-(1-\theta)W+\omega(1-\theta)(\mathbf{e}_1\mathbf{w}_1^\top+\mathbf{e}_j\mathbf{w}_j^\top), \]
	where $\mathbf{w}_1, \mathbf{w}_j$ are the first and $j$th column of $W$, respectively. Moreover, the optimization problem \eqref{eq:op_follower} becomes
	\begin{equation}\label{eq:op_cycle}
		\begin{aligned}
			&\mathrm{Maximize}_{j\in\V} \quad   \mathbf{1}^\top H(\s)^{-1}(\mathbf{e}_1+\mathbf{e}_j)  \\
			&\quad\mathrm{s.t.} \quad \quad j\not= 1
		\end{aligned}
	\end{equation}
	Note that $M:=I-(1-\theta)W$ is a strictly diagonally dominant circulant matrix, thus non-singular. According to the Sherman-Morrison formnula, it holds
	\begin{equation*}
		\begin{aligned}
			H(\s)^{-1}&=M^{-1}-\left.\omega(1-\theta)M^{-1}[\mathbf{e}_1,\mathbf{e}_j]\cdot\right(I+\\
			&\left.\omega(1-\theta)\left[\begin{array}{c}
				\mathbf{w}_1^\top  \\
				\mathbf{w}_j^\top
			\end{array}\right]M^{-1}[\mathbf{e}_1,\mathbf{e}_j]\right)^{-1}\left[\begin{array}{c}
				\mathbf{w}_1^\top  \\
				\mathbf{w}_j^\top
			\end{array}\right]M^{-1}
		\end{aligned}
	\end{equation*}
	From conclusions in literature \cite{searle1979inverting}, $M^{-1}$ is a circulant matrix. Let 
	\[M^{-1}=C(m_0,m_1,\dots,m_{n-1}). \]
	It then holds 
	\[\mathbf{1}^\top M^{-1}\mathbf{e}_1=\mathbf{1}^\top M^{-1}\mathbf{e}_j=\sum_{\ell=0}^{n-1}m_\ell, \]
	which is irrelevant of the choice of $\mathcal{S}$.
	Therefore, the optimization problem \eqref{eq:op_cycle} is equivalent to the following problem
	\begin{equation}\label{eq:op_cycle2}
		\begin{aligned}
			&\mathrm{Minimize}_{j\in\V} \quad   \mathbf{1}^\top \left(I+
			\omega(1-\theta)\left[\begin{array}{c}
				\mathbf{w}_1^\top  \\
				\mathbf{w}_j^\top
			\end{array}\right]M^{-1}[\mathbf{e}_1,\mathbf{e}_j]\right)^{-1}\\
			&\qquad\qquad\qquad\cdot\omega(1-\theta)\left[\begin{array}{c}
				\mathbf{w}_1^\top  \\
				\mathbf{w}_j^\top
			\end{array}\right]M^{-1}[\mathbf{e}_1,\mathbf{e}_j]\mathbf{1}   \\
			&\quad\mathrm{s.t.} \quad \quad j\not= 1
		\end{aligned}
	\end{equation}
	Define  
	\[B_j:=\omega(1-\theta)\left[\begin{array}{c}
		\mathbf{w}_1^\top  \\
		\mathbf{w}_j^\top
	\end{array}\right]M^{-1}[\mathbf{e}_1,\mathbf{e}_j]\in\mathbb{R}^{2\times 2}. \]
	Note that $(I+B_j)^{-1}B_j=I-(I+B_j)^{-1}$. The optimization problem \eqref{eq:op_cycle2} is then equivalent to
	\begin{equation}\label{eq:op_cycle3}
		\begin{aligned}
			&\mathrm{Maximize}_{j\in\V} \quad   \mathbf{1}^\top (I+B_j)^{-1}\mathbf{1}   \\
			&\quad\mathrm{s.t.} \quad \quad j\not= 1
		\end{aligned}
	\end{equation}
	Due to the fact that $(I-(1-\theta)W)M^{-1}=I$, i.e., $WM^{-1}=\frac{1}{1-\theta}(M^{-1}-I)$, it must be 
	\begin{equation*}
		\begin{aligned}
			\left[\begin{array}{c}
				\mathbf{w}_1^\top  \\
				\mathbf{w}_j^\top
			\end{array}\right]M^{-1}[\mathbf{e}_1,\mathbf{e}_j]&=\frac{1}{1-\theta}\left\{\left[\begin{matrix}
				[M^{-1}]_{11} &  [M^{-1}]_{1j} \\
				[M^{-1}]_{j1} & [M^{-1}]_{jj}
			\end{matrix}\right]-I\right\}\\
			&=\frac{1}{1-\theta}\left[\begin{matrix}
				m_0-1 &  m_{j-1} \\
				m_{1-j} & m_0-1
			\end{matrix}\right]
		\end{aligned}
	\end{equation*}
	Therefore,
	\[B_j=\omega\left[\begin{array}{cc}
		m_0-1 & m_{j-1}  \\
		m_{1-j} & m_0-1
	\end{array}\right]. \]
	Note that the matrix $M$ is symmetric. Therefore, $M^{-1}$ is also symmetric, which gives that $m_\ell=m_{n-\ell}, \forall \ell\in\{0,1,\dots, n\}$, and then $m_{1-j}=m_{j-1}, \forall j\in [n]$. Moreover, for the reason that 
	$M^{-1}=\sum_{k=0}^\infty (1-\theta)^kW^k$, it is $m_\ell>0, \ell=0,1,\dots, n-1$.
	As a result,
	\begin{equation*}
		\begin{aligned}
			\mathbf{1}^\top (I+B_j)^{-1}\mathbf{1} &=\frac{2\left(1+\omega (m_0-1)-\omega m_{j-1}\right)}{\left(1+\omega(m_0-1)\right)^2-\left(\omega m_{j-1}\right)^2}\\
			&= \frac{2}{1+\omega(m_0+m_{j-1}-1)}>0
		\end{aligned}
	\end{equation*}
	The optimization problem \eqref{eq:op_cycle3} then becomes
	\begin{equation}\label{eq:op_cycle4}
		\begin{aligned}
			&\mathrm{Minimize}_{j\in\V} \;\;   m_{j-1} \quad\mathrm{s.t.} \;\; j\not= 1
		\end{aligned}
	\end{equation}
	The proof is then completed.\hfill$\square$
	\\ \hfill \\
	
	To prove Theorem \ref{th:cycle}, some properties on product of symmetric circulant matrices are required. To make the main proof more clear, we only introduce the lemmas here, and leave their proofs to the next subsection. 
	
	\begin{lemma}\label{le:monotone}
		Given two real number sequences $u_0,u_1,\dots,u_{s-1}$ and $v_0,v_1,\dots,v_{s-1}$, such that $0\leq u_{h+1}\leq u_h$ and $0\leq u_{h+1}\leq u_h$ hold for all $h\in\{0,1,\dots,s-2\}$. The following statements then hold.
		\begin{enumerate}
			\item Define $U=C(u_0,u_1,\dots,u_{s-1},u_{s-1},\dots,u_1)$ and $V=C(v_0,v_1,\dots,v_{s-1},v_{s-1},\dots,v_1)$. Let $n=2s-1$ $R=UV\in\mathbb{R}^{n\times n}$. The matrix $R$ is a circulant matrix, denoted as $R=C(r_0,r_1,\dots,r_{s-1},r_{s-1},\dots,r_1)$. It holds $0\leq r_{h+1}\leq r_h, \forall h=0,1,\dots, s-2$. 
			\item Define $U=C(u_0,u_1,\dots,u_{s-1},0,u_{s-1},\dots,u_1)$ and $V=C(v_0,v_1,\dots,v_{s-1},0,v_{s-1},\dots,v_1)$. Let $n=2s$ $R=UV\in\mathbb{R}^{n\times n}$. The matrix $R$ is a circulant matrix, denoted as $R=C(r_0,r_1,\dots,r_{s-1},r_s,r_{s-1},\dots,r_1)$. It holds $0\leq r_{h+1}\leq r_h, \forall h=0,1,\dots,s-1$.
		\end{enumerate}
	\end{lemma}
	
	\begin{lemma}\label{le:strict_monotone}
		Consider two real number sequences $u_0,u_1,\dots,u_{s-1}$ and $v_0,v_1,\dots,v_{s-1}$, with $u_0\geq u_1\geq \dots\geq u_{s-1}$ and $v_0\geq v_1\geq \dots\geq v_{s-1}$. The following statements then hold.
		\begin{enumerate}
			\item Define $U=C(u_0,u_1,\dots,u_{s-1},u_{s-1},\dots,u_1)$ and $V=C(v_0,v_1,\dots,v_{s-1},v_{s-1},\dots,v_1)$. Let $R=UV=C(r_0,r_1,\dots,r_{s-1},r_{s-1},\dots,r_1)$. Suppose that $u_i>u_{i+1}$, $v_j>v_{j+1}$ and $i\geq j$ hold for some $i,j=0,1,\dots,s-2$. Then, for $h^\ast=\min\{i+j,2s-3-i-j\}$, it holds that $r_{i-j}>r_{i-j+1}>\dots>r_{h^\ast}>r_{h^\ast+1}$.
			\item Define $U=C(u_0,u_1,\dots,u_{s-1},u_s,u_{s-1},\dots,u_1)$ and $V=C(v_0,v_1,\dots,v_{s-1},v_s,v_{s-1},\dots,v_1)$. Let $R=UV=C(r_0,r_1,\dots,r_{s-1},r_s,r_{s-1},\dots,r_1)$. Suppose that $u_i>u_{i+1}$, $v_j>v_{j+1}$ and $i\geq j$ hold for some $i,j=0,1,\dots,s-1$. Then, for $h^\ast=\min\{i+j,2s-1-i-j\}$, it holds that $r_{i-j}>r_{i-j+1}>\dots>r_{h^\ast}>r_{h^\ast+1}$. 
		\end{enumerate}
	\end{lemma}
	
	Basically, Lemmas \ref{le:monotone} and \ref{le:strict_monotone} are saying that for two symmetric circulant matrices with monotone patterns in their rows (i.e., $u_h\geq u_{h+1}$ and $v_h\geq v_{h+1}$, the monotonicity is preserved in their product. 
	\hfill\break
	
	\noindent \textbf{Proof of Theorem \ref{th:cycle}}. We only prove the conclusion 1), and 2) can be obtained similarly. Consider 
	\[M^{-1}=\sum_{k=0}^\infty (1-\theta)^kW^k. \]
	From Lemma \ref{le:monotone}, and by induction, $W^k$ is a circulant matrix of the form $C(m_{0}^{(k)}, m_{1}^{(k)},\dots, m_{s-1}^{(k)},m_{s-1}^{(k)},\dots, m_{1}^{(k)})$, with $m_{0}^{(k)}\geq m_{1}^{(k)}\geq \dots\geq m_{s-1}^{(k)}$. As $w_0>w_{s-1}$, there exists $h\in\{1,\dots,s-2\}$ such that $w_h>w_{h+1}$. Applying Lemma~\ref{le:strict_monotone} to $W^2$ (with $i=j=h$), we obtain $h^\ast=\min\{2h,2s-2h-3\}\geq 1$. Define $\bar{h}_1=\min\{2h,2s-2h-3\}$, then
	\[m_0^{(2)}>m_1^{(2)}>m_2^{(2)}>\dots>m_{\bar{h}_1}^{(2)}>m_{\bar{h}_1+1}^{(2)}.\] 
	Again applying Lemma \ref{le:strict_monotone} to $W^4$ (with $U=W^2, V=W^2, i=j=1,2,\dots,\bar{h}_1$), we obtain 
	\[m_0^{(4)}>m_1^{(4)}>m_2^{(4)}>\dots>m_{\bar{h}_2}^{(4)}>m_{\bar{h}_2+1}^{(4)},  \]
	where $\bar{h}_2=\max_{\ell\in\{0,\dots,\bar{h}_1\}}\min\{2\ell, 2s-2\ell-3\}$. This process can be iterated, and we can obtain
	\[m_0^{(2^k)}>m_1^{(2^k)}>m_2^{(2^k)}>\dots>m_{\bar{h}_k}^{(2^k)}>m_{\bar{h}_k+1}^{(2^k)},  \]
	with $\bar{h}_k$ iteratively defined as
	\[\bar{h}_{k+1}:=\max_{\ell\in\{0,\dots,\bar{h}_k\}}\min\{2\ell, 2s-2\ell-3\}. \]
	Note that $\bar{h}_{k+1}=2\bar{h}_{k}$ if $2\bar{h}_{k}\leq s-2$. Therefore, there exists $k'$ such that $\bar{h}_{k'}=s-2$ (when $s-2$ is even) or $\bar{h}_{k'}=s-3$ (when $s-2$ is odd), and for the latter case it is $\bar{h}_{k'+1}=2s-3-2\cdot \frac{s-1}{2}=s-2$. As a result, we either have 
	\[m_0^{(2^{k'})}>m_1^{(2^{k'})}>m_2^{(2^{k'})}>\dots>m_{s-2}^{(2^{k'})}>m_{s-1}^{(2^{k'})}  \]
	or 
	\[m_0^{(2^{k'+1})}>m_1^{(2^{k'+1})}>\dots>m_{s-2}^{(2^{k'+1})}>m_{s-1}^{(2^{k'+1})}.  \]
	Note that $m_\ell =\sum_{k=0}^\infty m_{\ell}^{(k)}, \ell\in\{0,1,\dots, s-1\}$. Therefore, for both cases, it must be $m_0>m_1>\dots>m_{s-1}$. The proof is then completed.\hfill$\square$

	\section{Proofs of Lemmas \ref{le:monotone} and \ref{le:strict_monotone}}
	
	\noindent\textbf{Proof of Lemma \ref{le:monotone}}. Due to the fact that circulant matrices commute under multiplication \cite{searle1979inverting} and $U,V$ are both symmetric matrices, $R$ is also symmetric. Moreover, from the statement (iv) in \cite{searle1979inverting}, $R$ is a circulant matrix.
	\begin{enumerate}
		\item Let $\mathbf{u}=(u_0,u_1,\dots,u_{s-1},u_{s-1},\dots,u_1)^\top$, $\mathbf{v}=(v_0,v_1,\dots,v_{s-1},v_{s-1},\dots,v_1)^\top$, and $\mathbf{r}=(r_0,r_1,\dots,\allowbreak r_{s-1},r_{s-1},\dots,r_1)^\top$. We use $u(\ell), v(\ell)$ and $r(\ell)$ to denote the $(\ell+1)$th element of $\mathbf{u}, \mathbf{v}$ and $\mathbf{r}$, respectively. Again from the statement (iv) of \cite{searle1979inverting}, it holds $r(h)=\sum_{\ell=0}^{n-1}u(\ell) v(n+h-\ell),\forall h=0,1,\dots, n-1$. Here we make a convention that $u(n+h)=u(h), v(n+h)=v(h), \forall h\in\mathbb{Z}$. Note that $u(\ell)=u(-\ell), v(\ell)=v(-\ell), \forall \ell\in\mathbb{Z}$. We then have 
		\[r(h)=\sum_{\ell=0}^{n-1}u(\ell) v(h-\ell)=\sum_{\ell=0}^{n-1}u(\ell) v(\ell-h)  \]
		for all $h=0,1,\dots, n-1$. As a result, for $h=0,1,\dots, s-2$, it holds
		\begin{equation}
			\begin{aligned}
				r_h&=u_0v_{-h}+u_1v_{1-h}+\dots+u_hv_0\\
				&+u_{h+1}v_1+\dots+u_{s-1}v_{s-h-1}\\
				&+u_{-1}v_{-1-h}+\dots+u_{2-s+h}v_{2-s}+u_{1-s+h}v_{1-s}\\
				&+u_{-s+h}v_{-s}+\dots+u_{1-s}v_{1-s-h}\\
				&=u_0v_{h}+u_1v_{h-1}+\dots+u_hv_0\\
				&+u_{h+1}v_1+\dots+u_{s-1}v_{s-h-1}\\
				&+u_{1}v_{h+1}+\dots+u_{s-h-2}v_{s-2}+u_{s-h-1}v_{s-1}\\
				&+u_{s-h}v_{s-1}+\dots+u_{s-1}v_{s-h},\\
				r_{h+1}&=u_0v_{h+1}+u_1v_h+\dots+u_hv_1\\
				&+u_{h+1}v_0+\dots+u_{s-1}v_{s-h-2}\\
				&+u_1v_{h+2}+\dots+u_{s-h-2}v_{s-1}+u_{s-h-1}v_{s-1}\\
				&+u_{s-h}v_{s-2}+\dots+u_{s-1}v_{s-h-1}
			\end{aligned}
		\end{equation}
		\noindent\emph{Case 1: $h+1=s-h-1$}. We have 
		\begin{equation}\label{eq:sorting_ineq1}
			\begin{aligned}
				&r_h-r_{h+1}\\
				&=\sum_{\ell=1}^{s-h-1}\Big\{u_{h-\ell+1}(v_{\ell-1}-v_{\ell})+u_{h+\ell}(v_{\ell}-v_{\ell-1})\Big\}\\
				&+\sum_{\ell=1}^{s-h-2}\Big\{u_{s-h-\ell-1}(v_{s-\ell-1}-v_{s-\ell})\\
				&\qquad \qquad+u_{s-h+\ell-1}(v_{s-\ell}-v_{s-\ell-1})\Big\}\\
				&=\sum_{\ell=1}^{h+1}(u_{h-\ell+1}-u_{h+\ell})(v_{\ell-1}-v_{\ell})+\\
				&\sum_{\ell=1}^{s-h-2}(u_{s-h-\ell-1}-u_{s-h+\ell-1})(v_{s-\ell-1}-v_{s-\ell})\geq 0\\
			\end{aligned}
		\end{equation}
		\noindent\emph{Case 2: $h+1>s-h-1$}. We have 
		\begin{equation}\label{eq:sorting_ineq2}
			\begin{aligned}
				&r_h-r_{h+1}= \sum_{\ell=1}^{s-h-1}(u_{h-\ell+1}-u_{h+\ell})(v_{\ell-1}-v_{\ell})+\\
				&\sum_{\ell=1}^{s-h-2}(u_{s-h-1-\ell}-u_{s-h-1+\ell})(v_{s-1-\ell}-v_{s-1-\ell+1})\\
				&+u_0v_h+u_1v_{h-1}+\dots +u_{2h-s+1}v_{s-h-1}\\
				&-u_0v_{h+1}-u_1v_h-\dots-u_{2h-s+1}v_{s-h}\\
				&+u_{2s-2h-2}v_{h+1}+u_{2s-2h-1}v_{h}+\dots+u_{s-1}v_{s-h}\\
				&-u_{2s-2h-2}v_{h}-u_{2s-2h-1}v_{h-1}-\dots-u_{s-1}v_{s-h-1}\\
				&=\sum_{\ell=1}^{s-h-1}(u_{h-\ell+1}-u_{h+\ell})(v_{\ell-1}-v_{\ell})+\\
				&\sum_{\ell=1}^{s-h-2}(u_{s-h-1-\ell}-u_{s-h-1+\ell})(v_{s-1-\ell}-v_{s-1-\ell+1})\\
				&+ \sum_{\ell=0}^{2h-s+1}(u_{\ell}-u_{2s-2h-2+\ell})(v_{h-\ell}-v_{h-\ell+1})\geq 0.
			\end{aligned}
		\end{equation}
		\noindent\emph{Case 3: $h+1<s-h-1$}. Similarly to \eqref{eq:sorting_ineq2}, we have
		\begin{equation}\label{eq:sorting_ineq3}
			\begin{aligned}
				&r_h-r_{h+1}=\sum_{\ell=1}^{h+1}(u_{h-\ell+1}-u_{h+\ell})(v_{\ell-1}-v_{\ell})+\\
				&\sum_{\ell=1}^{h}(u_{s-h-1-\ell}-u_{s-h+1-\ell})(v_{s-\ell-1}-v_{s-\ell})\\
				&+ u_{2h+2}v_{h+2}+u_{2h+3}v_{h+3}+\dots+u_{s-1}v_{s-h-1}\\
				&-u_{2h+2}v_{h+1}-u_{2h+3}v_{2h+2}-\dots-u_{s-1}v_{s-h-2}\\
				&+u_1v_{h+1}+u_2v_{h+2}+\dots+u_{s-2h-2}v_{s-h-2}\\
				&-u_1v_{h+2}-u_2v_{h+3}-\dots-u_{s-2h-2}v_{s-h-1}\\
				&=\sum_{\ell=1}^{h+1}(u_{h-\ell+1}-u_{h+\ell})(v_{\ell-1}-v_{\ell})+\\
				&\sum_{\ell=1}^{h}(u_{s-h-1-\ell}-u_{s-h+1-\ell})(v_{s-\ell-1}-v_{s-\ell})\\
				&+\sum_{\ell=1}^{s-2h-2}(u_{\ell}-u_{2h+1+\ell})(v_{h+\ell}-v_{h+\ell+1})\geq 0.
			\end{aligned}
		\end{equation}
		
		
		Combining all the arguments above, the conclusion is proved. 
		\item The proof is similar to that of 1), thus omitted. \hfill$\square$
	\end{enumerate}
	\hfill\break
	
	\noindent\textbf{Proof of Lemma \ref{le:strict_monotone}}. We only prove 1), and 2) can be proved similarly. 
	
	\emph{Step 1}. Let
	\begin{equation*}
		\begin{aligned}
			\bar{U}&:=C(u_i\mathbf{1}_{i+1}^\top, u_{i+1}\mathbf{1}_{s-1-i}^\top,u_{i+1}\mathbf{1}_{s-1-i}^\top,u_i\mathbf{1}_{i}^\top),\\
			\bar{V}&:=C(v_i\mathbf{1}_{i+1}^\top, v_{i+1}\mathbf{1}_{s-1-i}^\top,v_{i+1}\mathbf{1}_{s-1-i}^\top,v_i\mathbf{1}_{i}^\top),\\
			\bar{R}&:=\bar{U}\bar{V}=C(\bar{r}_0,\bar{r}_1,\dots,\bar{r}_{s-1},\bar{r}_{s-1},\dots, \bar{r}_1).
		\end{aligned}
	\end{equation*}
	We prove that if $\bar{r}_h>\bar{r}_{h+1}$ for some $h=0,1,\dots,s-2$, it must be $r_h>r_{h+1}$. From \eqref{eq:sorting_ineq1}-\eqref{eq:sorting_ineq3}, it can be seen that if $\bar{r}_h>\bar{r}_{h+1}$, at least one term in the summation (corresponding to the value of $h$) is strictly positive, say, $h+1>s-h-1$ and $(\bar{u}_{h-\ell+1}-\bar{u}_{h+\ell})(\bar{v}_{\ell-1}-\bar{v}_\ell)>0$ hold for some $\ell=1,\dots,s-h-1$ (here we let $(\bar{u}_0,\bar{u}_1,\dots,\bar{u}_{s-1})=(u_i\mathbf{1}_{i+1}^\top,u_{i+1}\mathbf{1}_{s-1-i}^\top)$ and $(\bar{v}_0,\bar{v}_1,\dots,\bar{v}_{s-1})=(v_i\mathbf{1}_{i+1}^\top,v_{i+1}\mathbf{1}_{s-1-i}^\top)$). Then, it must be $\ell=j+1$ and $0\leq h-\ell+1\leq i<h+\ell\leq s-1$, which further gives $u_{h-\ell+1}\geq u_{i}> u_{i+1}= u_{h+\ell}$ and $v_{\ell-1}=v_{j}>v_{j+1}=v_\ell$. Again from \eqref{eq:sorting_ineq2}, we obtain $r_h>r_{h+1}$. The argument also holds if \eqref{eq:sorting_ineq1} or \eqref{eq:sorting_ineq3} holds for $h$. The claim is then proved.
	
	\emph{Step 2}. Let $u:=u_{i}-u_{i+1}>0$, $v:=v_i-v_{i+1}>0$, and 
	\begin{equation*}
		\begin{aligned}
			\tilde{U}&=C(u\mathbf{1}_{i+1}^\top,\mathbf{0}_{s-1-i}^\top,\mathbf{0}_{s-1-i}^\top,u\mathbf{1}_{i}^\top ),\\
			\tilde{V}&=C(v\mathbf{1}_{i+1}^\top,\mathbf{0}_{s-1-i}^\top,\mathbf{0}_{s-1-i}^\top,v\mathbf{1}_{i}^\top ),\\
			\tilde{R}&:=\tilde{U}\tilde{V}=C(\tilde{r}_0,\tilde{r}_1,\dots,\tilde{r}_{s-1},\tilde{r}_{s-1},\dots, \tilde{r}_1).
		\end{aligned}
	\end{equation*}
	We prove that if $\tilde{r}_h>\tilde{r}_{h+1}$ holds for some $h=0,1,\dots,s-2$, it must be $\bar{r}_h>\bar{r}_{h+1}$. Note that $\tilde{U}=\bar{U}-u_{i+1}\mathbf{1}\mathbf{1}^\top$ and $\tilde{V}=\bar{V}-v_{i+1}\mathbf{1}\mathbf{1}^\top$. We then have $\tilde{R}=\bar{R}-n(u_{i+1}+v_{i+1}-u_{i+1}v_{i+1})\mathbf{11}^\top$. Therefore, $\tilde{r}_h>\tilde{r}_{h+1}$ is equivalent to $\bar{r}_h>\bar{r}_{h+1}$.
	
	\emph{Step 3}. Consider $\tilde{R}$. We prove that 
	\begin{equation}\label{eq:ineq_rtilde}
		\tilde{r}_{i-j}>\dots>\tilde{r}_{h^\ast}>\tilde{r}_{h^\ast+1}.
	\end{equation}
	Without loss of generality, let $j\leq i$. 
	\begin{itemize}
		\item If $i+j\leq s-2$, by calculation,
		\begin{equation*}
			\begin{aligned}
				&\tilde{r}_0=\tilde{r}_1=\dots=\tilde{r}_{i-j}=(2j+1)uv,\\ 
				&\tilde{r}_{i-j+1}=2juv,\dots,\tilde{r}_{i+j}=uv,\\
				&\tilde{r}_{i+j+1}=\tilde{r}_{i+j+2}=\dots=\tilde{r}_{s-1}=0,
			\end{aligned}
		\end{equation*}
		which gives \eqref{eq:ineq_rtilde}.
		\item If $i+j=s-1$, by calculation, 
		\begin{equation*}
			\begin{aligned}
				&\tilde{r}_0=\tilde{r}_1=\dots=\tilde{r}_{i-j}=(2j+1)uv,\\ 
				&\tilde{r}_{i-j+1}=2juv,\dots,\tilde{r}_{s-1}=uv,
			\end{aligned}
		\end{equation*}
		which gives \eqref{eq:ineq_rtilde}.
		\item If $i+j\geq s$, by calculation,
		\begin{equation*}
			\begin{aligned}
				&\tilde{r}_0=\tilde{r}_1=\dots=\tilde{r}_{i-j}=(2j+1)uv,\\ 
				&\tilde{r}_{i-j+1}=2juv,\dots,\tilde{r}_{2s-3-i-j}=(2i+2j-2s+4)uv,\\
				&\tilde{r}_{2s-2-i-j}=\dots=\tilde{r}_{s-1}=(2i+2j-2s+3)uv,
			\end{aligned}
		\end{equation*}
		which again gives \eqref{eq:ineq_rtilde}.
	\end{itemize}
	Combining all the arguments above, the desired conclusion is obtained. \hfill$\square$


\end{document}